\documentclass[pdflatex,sn-mathphys-num]{sn-jnl}

\usepackage{graphicx}%
\usepackage{multirow}%
\usepackage{amsmath,amssymb,amsfonts}%
\usepackage{amsthm}%
\usepackage{mathrsfs}%
\usepackage[title]{appendix}%
\usepackage{xcolor}%
\usepackage{textcomp}%
\usepackage{manyfoot}%
\usepackage{booktabs}%
\usepackage{algorithm}%
\usepackage{algorithmicx}%
\usepackage{algpseudocode}%
\usepackage{listings}%

\usepackage{hyperref}   
\hypersetup{
	colorlinks=true,
	linkcolor=blue,
	citecolor=blue,
	urlcolor=blue,
}

\usepackage{mathtools}

\usepackage{lipsum}
\usepackage{diagbox}
\usepackage{indentfirst}

\usepackage{tabularray}
\usepackage{tabularx}
\usepackage{multirow}
\usepackage{geometry}
\usepackage{hhline}
\usepackage{makecell}
\DeclarePairedDelimiterX\set[1]\{\}{\nonscript\,#1\nonscript\,}
\usepackage{cleveref}
\usepackage[parfill]{parskip}
\usepackage{threeparttable}
\usepackage[numbers]{natbib}
\usepackage{bm}
\usepackage{amsmath}   
\usepackage{kotex}

\theoremstyle{thmstyleone}%
\theoremstyle{thmstyletwo}%

\theoremstyle{thmstylethree}%

\begin{document}

\title[Article Title]{Modular Mechanism Design Optimization in Large-Scale Systems with Manufacturing Cost Considerations}


\author[1]{Sumin Lee}\email{smlee1996@kaist.ac.kr}

\author*[2,3]{Namwoo Kang}\email{nwkang@kaist.ac.kr}

\affil[1]{Department of Mechanical Engineering, Korea Advanced Institute of Science and Technology (KAIST), 291 Daehak-ro, Yuseong-gu, Daejeon 34141, Republic of Korea}

\affil[2]{Cho Chun Shik Graduate School of Mobility, Korea Advanced Institute of Science and Technology (KAIST), 193 Munji-ro, Yuseong-gu, Daejeon 34051, Republic of Korea}

\affil[3]{Narnia Labs, 193 Munji-ro, Yuseong-gu, Daejeon 34051, Republic of Korea}


\abstract{The modular design maximizes utility by employing standardized rather than customized components in large-scale systems. From a manufacturing view, this approach contributes to green technology by reducing unnecessary material waste and enhancing component reusability. Additionally, from an industrial view, it offers significant economic advantages by leveraging economies of scale, making it a highly reasonable design strategy. Modularization is generally achieved by selecting a representative design that satisfies all required performance from a predefined set of design candidates. However, achieving effective modularization in mechanical mechanism systems presents additional challenges. First, mechanical mechanisms rely on the geometric relationships between components to achieve functional motion, and variations in load conditions lead to different optimal design parameters. As a result, selecting a representative design for modularization is inherently complex. Second, the selected representative design often exceeds the optimal design parameters, leading to over-specification. This results in inevitable performance deviations, which become more pronounced as the design scale increases. To address these challenges, this study proposes a modular mechanism design framework based on surrogate-based design optimization. Surrogate-based optimization is utilized to obtain optimal design solutions for each component in large-scale engineering systems. The generated optimal designs are then partitioned into multiple groups, with an optimal design assigned to each group. This formulation is a multi-objective optimization (MOO) problem that aims to maximize economies of scale while minimizing performance deviations. Unlike conventional approaches, which primarily rely on predefined design candidates and simple grouping strategies, the proposed framework utilizes surrogate-based design optimization, allowing for more flexible design variable selection and modular optimization strategies. Additionally, this study analyzes the parameters defining the manufacturing cost curve, establishing a decision support system that enables the selection of optimal strategies across diverse design scenarios. This approach enhances maintainability, improves component interchangeability, and contributes to an environmentally sustainable manufacturing environment.}


\keywords{Modular design, large-scale design, design optimization, surrogate model}



\maketitle

\section{Introduction}\label{sec1}
Modular design is an essential strategy for achieving environmentally sustainable manufacturing, as it contributes to implementing sustainable production systems by extending product lifespan, enhancing maintainability, promoting component reuse, and minimizing waste generated during the manufacturing process~\cite{van2017design, ettlie2008design, elmaraghy2013product, hong2014modular, kimura2001product, yu2011product, tseng2023multi, kurniadi2020maintaining, wu2018design, wang2018optimization, lee2017remanufacturing, he2015product, kim2022usage}. This concept is not new and has been extensively studied and applied across various industries over time. In mechanical design and manufacturing, the interpretation of modularity varies depending on the application. Although defining modular design in a single, rigid framework is challenging, it is generally characterized as a strategy that maximizes flexibility in design and manufacturing processes while facilitating efficient system management~\cite{ramani2010integrated, moon2010methodology, kristjansson2004term, pandremenos2009modularity, pakkanen2019identifying, otto2016global, moon2010module, kim2017sustainable, moon2006data, kim2022usage, sinha2020design, sinha2018integrative}.

The adoption of modular design enables both resource efficiency and environmental sustainability. Standardization of components simplifies maintenance and replacement while maximizing reusability and recyclability~\cite{kimura2001product, pakkanen2019identifying, lee2017remanufacturing, kim2022usage, yao2016cost, suh2007flexible, suh2010technology}. In the long term, this approach is crucial in reducing waste throughout the production process. Furthermore, modularized systems streamline manufacturing by eliminating unnecessary production steps, minimizing resource waste, and supporting the realization of more environmentally friendly manufacturing practices. For these reasons, modular design has been widely recognized in the manufacturing industry as a fundamental design principle for achieving economic efficiency and environmental sustainability.

\begin{figure}[h!]
\centering
 \includegraphics[width=0.5\textwidth]{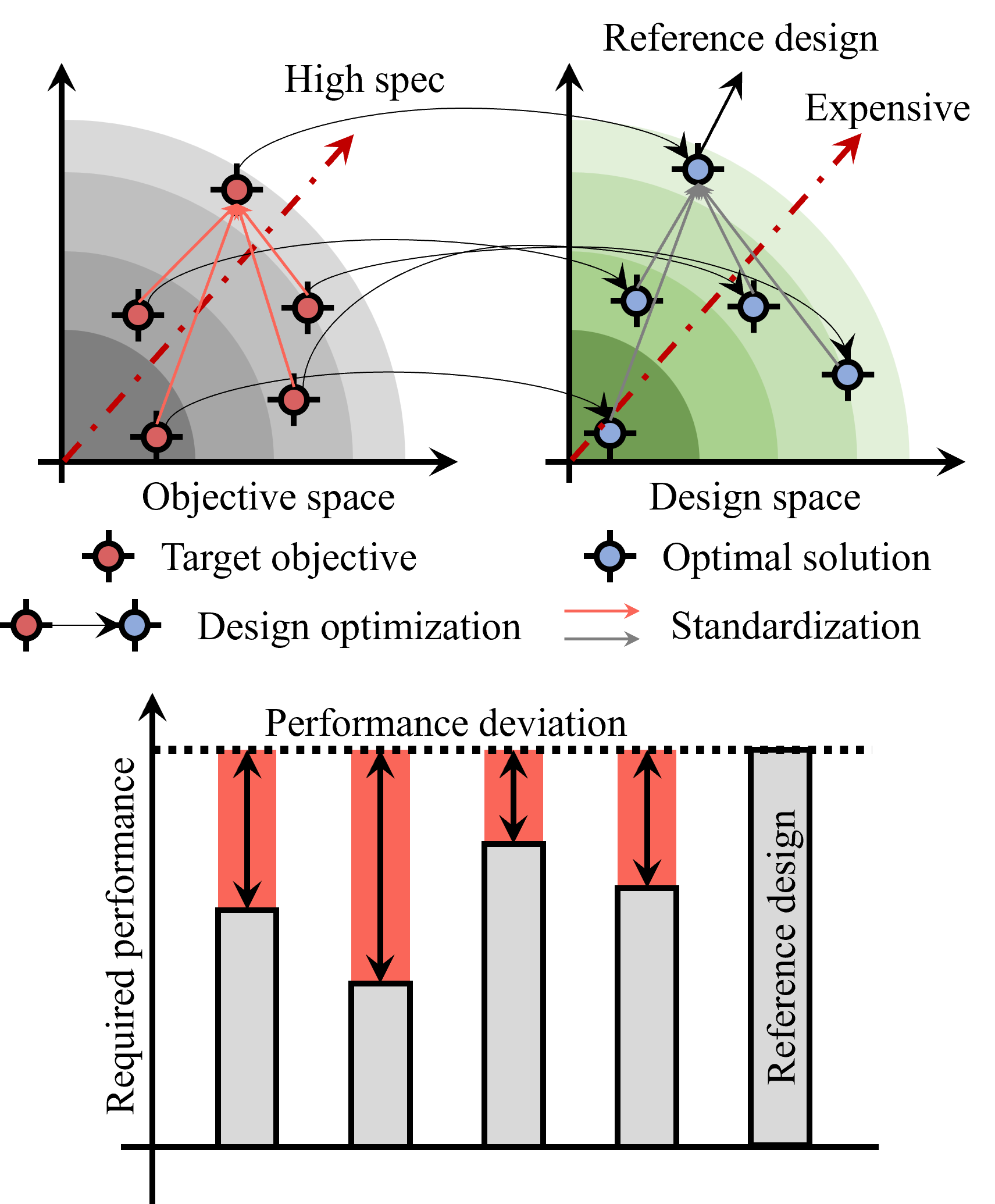}
\caption{Standardizing designs using high-spec solutions can lead to increased performance deviation.}
\label{fig_modular_performance_deviation}
\vspace{-9pt}
\end{figure}

Most manufacturing industries employ mass production methods to leverage economies of scale, with cost reduction being a primary objective~\cite{van2017design, pakkanen2019identifying, pandremenos2009modularity,hong2014modular, wu2018design, luh2020smart, da2001mass, kim2017correlation, sinha2018integrative, sinha2018pareto, sinha2020design, alptekinouglu2008mass, jung2023additive, zipkin2001limits, moon2006data}. In such production systems, standardized processes are repetitively applied to manufacture identical components quickly and consistently, maximizing production efficiency and minimizing raw material consumption. In this context, modular design is highly compatible with mass production systems, offering cost reduction and significant benefits in terms of environmental sustainability. Modular design simplifies manufacturing processes and reduces unnecessary material waste by enabling product standardization. Additionally, beyond the direct cost-saving benefits of mass production, modularization improves maintainability and enhances overall resource efficiency.
However, achieving a practical modular design requires addressing several critical considerations. First, it is essential to establish compatibility among system components and select representative design configurations that can be widely utilized. Failure to ensure sufficient engineering performance in modularized designs may lead to performance degradation across the entire system~\cite{mourtzis2020simulation, yoshimura2007system, zipkin2001limits}. Second, a strategy is required to minimize performance deviations from integrating standardized design configurations, as shown in Fig.~\ref{fig_modular_performance_deviation}. Simply applying a single design that exceeds the required load conditions can lead to unnecessary over-specification, incorporating excessive design elements, and increasing costs. This over-specification issue can result in higher material consumption and increased energy usage, ultimately conflicting with environmentally sustainable design principles. Therefore, a strategic approach is necessary to balance performance requirements and resource optimization within the modularization process.

\subsection{General engineering system design optimization}
\begin{figure}[thb]
\centering
 \includegraphics[width=0.85\textwidth]{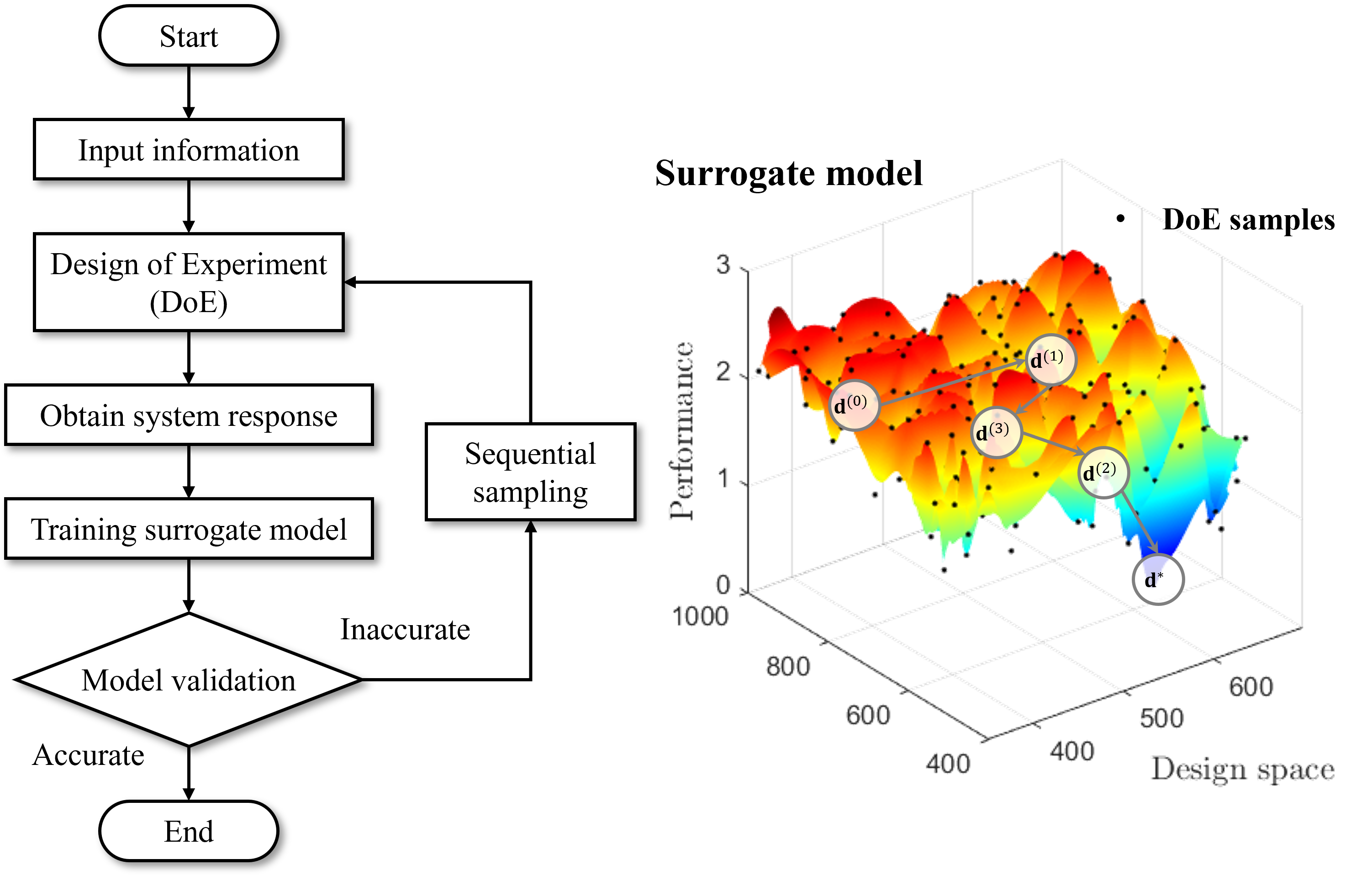}
\caption{Surrogate-based optimization process.}
\label{fig_surrogate_model}
\vspace{-9pt}
\end{figure}

Engineering systems are purposefully designed to perform intended functions. They are composed of multiple subsystems, each interacting with one another to achieve the desired functionality~\cite{ gokpinar2010impact, van2017design, luccarelli2015modular, park2015assessment, park2019investigation, suh2010technology, chiriac2011level, holtta2012comparative, kim2017correlation}. While each subsystem is designed to implement partial functionalities, the overall system can only fulfill its intended purpose when these components are organically integrated. This characteristic of engineering design is not limited to individual products or devices but applies to large-scale processes and automated manufacturing systems.
Understanding the relationship between the design space and the objective space is essential to ensuring the engineering performance required to achieve functional objectives. In traditional design methodologies, this relationship is determined via numerous iterations, where decisions are made based on whether a given design satisfies thresholds~\cite{kumagai2023black, wu2020knowledge, lambora2019genetic, mirjalili2019genetic, sobol2001global}. However, this approach entails significant computational costs for each iteration. It has inherent limitations, as it may fail to converge for problems where the potential of the design space remains unexplored. To address these challenges more effectively, surrogate-based optimization has recently emerged as a promising alternative.
This approach trains a mapping that predicts the objective space from the design space, generally following the standard process shown in Fig.~\ref{fig_surrogate_model}. This process typically consists of several key stages, including design of experiments (DoE), engineering analysis, surrogate modeling, and design optimization. Once a well-trained surrogate model is available, additional iterations can be performed without incurring significant computational costs, enabling a rapid assessment of the potential within the objective space. Consequently, this method is highly effective for design exploration and optimization.
However, in large-scale design scenarios where multiple types must be developed within the same product family, differences in target performance are frequently encountered. When each component of an engineering system is individually optimized, the resulting design variables often differ significantly across components, even when they serve identical functions~\cite{zawadzki2016smart, herrmann2014sustainability, da2001mass, moon2010module}. This variability becomes more pronounced as the number of components requiring optimization for distinct load conditions increases. Such inconsistencies fundamentally conflict with the prevailing industrial practice of leveraging economies of scale in mass production to achieve cost reductions~\cite{krahe2022ai, dahlgren2013small, van2017design, elmaraghy2013product, radder1999mass, jung2023additive}. As a result, they can lead to inefficiencies in manufacturing processes and increased production costs. 
Thus, considering both the implementation of environmentally sustainable technologies and economic feasibility, individually optimized designs may not always provide a reasonable solution in large-scale design scenarios~\cite{thomassen2017mass, krahe2022ai, renjith2020design, park2015assessment, yao2016cost, suh2007flexible_2, sinha2018pareto, zipkin2001limits}. A single optimal design is no longer sufficient to meet the demands of modern industries, particularly in large-scale automated manufacturing systems such as smart factories. These systems require integrating numerous facilities and production stages to maximize overall productivity. Even when individual design optimization is applied, a lack of consistency across different equipment performing distinct tasks can lead to inefficiencies in the manufacturing process. Therefore, it is essential to adopt design strategies that ensure system-wide consistency beyond optimizing individual designs. Nevertheless, existing design methodologies, including generative design approaches, primarily focus on optimizing or generating individual design candidates. As a result, ensuring design commonality within the same product family remains a significant challenge. This limitation constrains the cost-efficient development of large-scale engineering systems and highlights the necessity for modular design and standardization strategies to enhance design efficiency~\cite{yoshimura2007system, mourtzis2020simulation, luccarelli2015modular, guo2019modular, zuehlke2010smartfactory, radder1999mass, koren2015product, sakundarini2015methodology, luh2020smart, wright1936factors, da2001mass}. From a design engineering perspective, large-scale design scenarios should offer a diverse set of design alternatives but also ensure that these alternatives meet both engineering performance requirements and manufacturability constraints. While various design solutions may be required to accommodate different operational conditions and engineering demands, relying solely on individually optimized designs poses limitations when considering manufacturing cost efficiency.



\subsection{Problem definition}
This study proposes a large-scale industrial robot design scenario in which robots within the same product family are required to perform different tasks. In large-scale manufacturing conditions, each task demands specific optimal design variables and corresponding payload torque for industrial robots. The proposed industrial robot is implemented using a quasi-serial mechanism. To address these challenges, Lee et al. proposed a framework that considers both the geometric and dynamic conditions of robot design variables~\cite{lee2024multi}. This framework employs a surrogate model to predict and optimize the mechanism design variables of quasi-serial manipulators along with the associated payload torque. Doing so provides optimal design solutions tailored to input task requirements, recommending the most suitable design variables and corresponding payload torque. While this study does not delve into the detailed workings of the framework, the overall process follows four main steps: 1) parametric design, 2) engineering analysis, 3) surrogate modeling, and 4) design optimization. Through this systematic approach, the framework derives optimal design variables and payload torque based on the given task region and payload requirements.

Despite the effectiveness of this framework in optimizing design variables for specific tasks, several critical challenges emerge in large-scale design scenarios. First, design complexity increases exponentially when numerous design solutions are individually optimized and manufactured. This complexity can lead to significant resource waste during production and maintenance processes, an issue that becomes more pronounced as the design scale expands. Second, simple grouping strategies may result in geometrically infeasible design solutions due to the strong geometric dependency among design variables in mechanical mechanisms. Such infeasibility obstructs practical assembly and performance realization during manufacturing and increases the risk of generating infeasible designs.

Previous studies have attempted to address these challenges through individual design optimization or simple grouping strategies~\cite{otto2016global, elmaraghy2012complexity, hong2014modular}. However, achieving cost reduction and design efficiency becomes significantly more complex in large-scale design scenarios where manipulators are composed of quasi-serial mechanisms rather than a single part. The intricate interdependencies among components introduce additional constraints, making balancing standardization with performance optimization difficult. 

To effectively address these challenges, a modular approach based on generative design techniques is required to ensure design adaptability while meeting engineering performance requirements. By integrating modular principles with generative methodologies, this approach not only mitigates the inefficiencies of conventional methods but also enables scalable and standardized design solutions suitable for large-scale applications.
Building on this foundation, this study extends an individually optimized design framework to accommodate large-scale design scenarios. Instead of merely compiling a set of individually optimized designs, the proposed method systematically explores strategies for constructing an optimal modular layout that balances performance consistency and manufacturing cost within an economy of scale.

\subsection{Research goal}

This study proposes a modular design methodology for efficiently structuring large-scale design candidates in industrial applications. Unlike traditional design approaches, which focus solely on maximizing performance through individual optimization, the proposed modular design strategy systematically organizes designs based on a surrogate-based optimization framework. Ultimately, the proposed framework is not limited to large-scale manipulator design but can be extended to a wide range of applications requiring modular design solutions based on surrogate models. By applying this approach across different engineering domains, the framework enhances scalability and systematically optimizes modular structures in various industrial contexts.

The proposed modularization strategy aims to maximize economies of scale by effectively grouping design candidates while addressing the inherent trade-offs between group-based standardization and performance deviation. The optimization approach minimizes overspecification by accurately capturing the performance variance from the grouping process. It also considers the cost curves and parameters associated with various manufacturing applications, ensuring cost efficiency. By balancing these trade-offs, the strategy maintains economic viability and performance consistency across different production scenarios.

To achieve this, the study formulates cost functions that include manufacturing cost parameters, providing objective criteria for optimizing modular designs in practical applications. Rather than aggregating individually optimized solutions, the proposed approach integrates generative design methodologies with modular optimization principles to systematically enhance productivity and cost efficiency. This strategy enables flexible adaptation to evolving manufacturing environments and market demands by leveraging generative design models. The research focuses on three key objectives:

\begin{enumerate}
    \item \textbf{Balancing design commonality and engineering performance} \\
    Develop a quantitative framework to maintain design commonality while accommodating diverse performance requirements. This objective minimizes design complexity without compromising performance, leading to a cost-effective design strategy.\\

    \item \textbf{Optimizing trade-offs between manufacturing costs and performance deviation} \\
    Establish optimized design groups that reduce complexity, minimize manufacturing costs, and meet task-specific performance requirements. The approach aims to leverage economies of scale while avoiding excessive (overspec) or insufficient (underspec) performance specifications.\\

    \item \textbf{Developing a scalable modular design strategy} \\
    Create a scalable design strategy that adapts to advancements in manufacturing technologies and evolving market demands. The proposed method employs generative design models to effectively implement modularization, ensuring an optimal balance between cost and performance in mass production environments.

\end{enumerate}

The structure of this paper is as follows: Section~\ref{sec2} provides a summary of relevant studies on modular design. Section~\ref{sec3} presents an overview of the proposed framework and details each stage of the methodology. Section~\ref{sec4} discusses the results of the framework and analyzes the impact of key parameters on the objective function used to determine manufacturing costs. Finally, Section~\ref{sec5} concludes the paper by summarizing key findings, discussing limitations, and outlining directions for future work.

\section{Related works}\label{sec2}
In the mechanical engineering domain, modular design is typically implemented by structuring systems into independent modules, allowing each module to be designed, manufactured, and assembled separately to meet various performance requirements. A representative example is computer systems, where components such as the CPU, memory, and storage devices are designed for easy replacement and customization. However, excessive modularization can lead to performance degradation and increased design costs, making balancing performance requirements and cost efficiency crucial. Standardization and grouping strategies are commonly employed to address these challenges and maximize the advantages of modularity~\cite{zawadzki2016smart, koren2015product, sakundarini2015methodology, luh2020smart, krahe2022ai, perera1999component, antoniolli2017standardization, holtta2012comparative, park2015assessment, yao2016cost, moon2006data, moon2010module, ko2015design, otto2016global, suh2007flexible, suh2007flexible_2, kim2016analysis}.

Standardization is fundamental to reducing design complexity and improving production efficiency by defining and setting common design elements or components that can be utilized across various product families. Representative examples include mechanical elements such as bolts, bearings, and standard electronic components, which facilitate maintenance and upgrades, extending product lifespan while reducing costs~\cite{perera1999component, kimura2001product, pakkanen2019identifying, pandremenos2009modularity, elmaraghy2013product, park2015assessment, moon2014platform, moon2006data, kim2017sustainable, otto2016global, suh2007flexible, chiriac2011level, sinha2018integrative, jung2020domain}.
Meanwhile, grouping is a strategy aimed at achieving economies of scale by consolidating designs with similar performance requirements into a limited number of groups. Instead of optimizing each design individually, representative designs that satisfy a given range of performance requirements are identified, enabling the formation of diverse product families~\cite{kimura2001product, van2017design, pakkanen2019identifying}. A typical example is the manufacturing of medium- and heavy-duty trucks, where a standard chassis and engine platform is utilized, with specific components adjusted based on load capacity to create different models~\cite{kopp2012new, luccarelli2015modular, pandremenos2009modularity}. 
Such platform-based modularization strategies not only reduce design complexity and production costs but also enhance the optimization of product configurations and the flexibility of design modifications, allowing manufacturers to effectively meet diverse consumer demands~\cite{song2011research, yim2002modular, simpson2006platform}. A notable example is Ford Motor Company, which successfully optimized its production process while fulfilling customer needs by offering a modular selection system that enabled over 3.8 million vehicle configurations~\cite{simpson2006platform}.

Modularization strategies are a powerful approach for achieving economies of scale in mass production. However, sometimes they present more significant challenges than individual optimization due to the necessity of maintaining design uniformity~\cite{otto2016global, elmaraghy2012complexity, hong2014modular, pandremenos2009modularity, elmaraghy2013product, van2017design, krahe2022ai, dahlgren2013small, radder1999mass, jung2023additive}. Ultimately, modularization to leverage mass production systems can be achieved by unifying design elements with superior performance. This approach often involves standardizing components using high-spec designs derived from the objective space of previous surrogate-based optimization methods. However, as depicted in Fig.~\ref{fig_modular_performance_deviation}, this standardization can lead to trade-offs between minimizing performance deviation and maintaining cost efficiency. Specifically, while unification through high-spec designs may ensure performance consistency, it also introduces inefficiencies and higher costs due to the need for over-specification and the mismatch between the design and objective spaces. Consequently, various optimization approaches have been proposed to address this issue. 

One notable approach involves an adapted Taguchi methodology designed to enhance flexibility in responding to evolving customer demands. This method enables the prediction of future customer requirements while optimizing the relationship between quality characteristics and design variables~\cite{jiang2001design}. 
Modular design approaches considering the product lifecycle have been extensively studied~\cite{sakundarini2015methodology, he2015product, meng2015rapid, lee2017remanufacturing, perera1999component, yu2011product, martinez2016development, kimura2001product, elmaraghy2013product, ramani2010integrated}. Traditional studies have proposed methodologies that integrate functional attributes and lifecycle properties into modular design. This approach leverages Modular Driving Forces (MDFs) to optimize component interactions and employs the Group Genetic Algorithm (GGA) to derive optimal module configurations~\cite{yu2011product}. Similarly, adaptive optimization strategies have been developed to accommodate reconfigurations during operation and the development of next-generation models, utilizing lifecycle properties as a basis for module grouping~\cite{martinez2016development}. 
Furthermore, multi-objective optimization techniques have been employed to maximize the effectiveness of modularization strategies. In particular, Multi-objective Grouping Genetic Algorithms (MOGGAs) have demonstrated superior performance over conventional GGAs by simultaneously optimizing multiple objectives~\cite{tseng2023multi}.

In mechanical and manufacturing industries, the concept of product families has been introduced to facilitate the sharing of standard components and modules~\cite{yim2002modular, park2015assessment, moon2014platform, kim2017sustainable, moon2010module, suh2007flexible, kim2016analysis, kimura2001product, sinha2018pareto}. This approach enhances the efficiency of next-generation model development, upgrades, and derivative designs, making it an effective strategy adopted by various companies. For instance, Ford Motor Company successfully reduced development costs and maximized production efficiency by implementing a shared platform strategy across different vehicle models~\cite{simpson2006platform}. 
The product family concept is crucial in meeting diverse customer demands while increasing production flexibility and maintaining quality consistency in mass production systems. In the automotive industry, platform-based design enables multiple models to share the same chassis and powertrain, simplifying manufacturing processes, maximizing component reuse, and reducing costs while improving maintainability~\cite{kopp2012new, pandremenos2009modularity, holtta2012comparative, suh2020multi, kurniadi2020maintaining}. Similarly, research in the home appliance industry has focused on adopting standardized modules to enhance productivity and scalability for model upgrades. 
The application of modular design is also expanding in manufacturing systems and factory automation (smart factories). Studies have proposed modularizing factory equipment and robotic systems to enhance productivity and enable flexible responses to diverse production requirements~\cite{zuehlke2010smartfactory, guo2019modular, song2011research, yim2002modular, chu2016design, lee2024multi, dahlgren2013small, moon2006data, zawadzki2016smart}. In particular, industrial robot design has explored approaches that allow multiple tasks to be performed using the same modular components. This strategy facilitates adaptation to new work environments with minimal design modifications, emphasizing scalability and maintainability as key advantages~\cite{song2011research, yim2002modular}.

Existing studies have explored modular design strategies to maximize benefits such as reducing production costs~\cite{hong2014modular, chu2016design, wu2018design}, enhancing component reuse~\cite{kimura2001product, van2017design, pakkanen2019identifying, lee2017remanufacturing}, and improving maintainability~\cite{pandremenos2009modularity, elmaraghy2013product, mesa2019trends}. However, these studies commonly focus on managing large-scale production with predetermined design variables, emphasizing cost efficiency and environmental sustainability. While such approaches effectively optimize manufacturing processes under fixed design candidates, they inherently limit design flexibility by relying on predefined design solutions.

This study proposes a modular design strategy that leverages optimal design solutions generated directly from an optimization framework rather than relying on a predefined set of design candidates. Integrating a surrogate-based optimization model upstream enables the exploration of a large pool of optimized designs, providing a more flexible and scalable modular design framework. Considering cost efficiency and performance deviation, the proposed methodology enhances design standardization in large-scale manufacturing environments.

\newpage
\section{Modular mechanism design optimization framework}\label{sec3}
\begin{figure*}[thb]
\centering
 \includegraphics[width=1\textwidth]{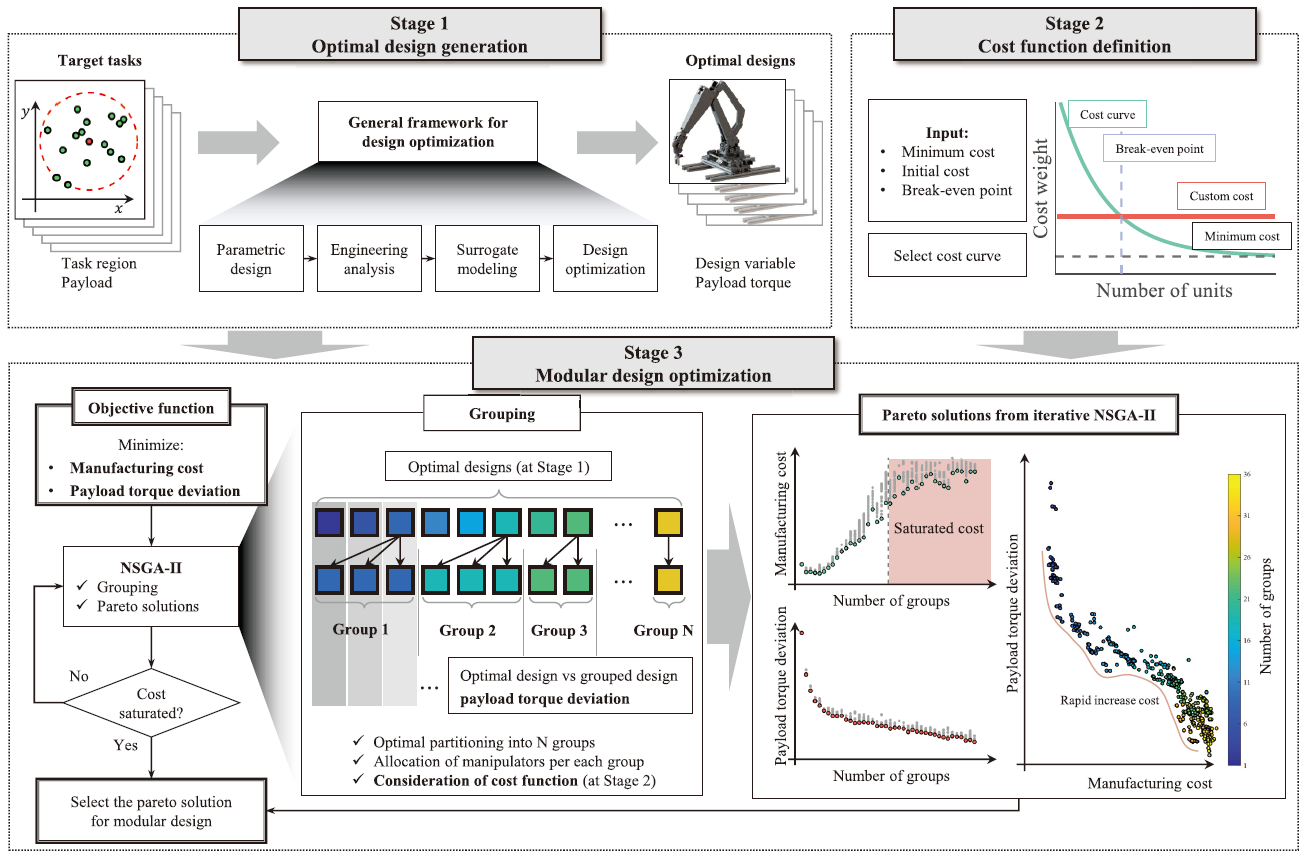}
\caption{Modular design optimization framework with a focus on manufacturing cost efficiency.}
\label{CH6_fig_framework}
\vspace{-9pt}
\end{figure*}

The proposed modular design framework in this study consists of three stages, as illustrated in Fig.~\ref{CH6_fig_framework}. Stage 1, the framework utilizes the optimization results of quasi-serial manipulators to generate optimized designs for diverse tasks. At this stage, manufacturing costs are not yet considered. Instead, the primary focus is mapping the optimal design variables and the required payload torque to each task, providing the most suitable manipulator design according to specific task requirements. Stage 2 involves defining manufacturing costs by balancing individual and mass production trade-offs. These costs are calibrated based on technological advancements and the specific application environment. To effectively implement the modular design strategy, a tailored cost curve is established to assign appropriate weight factors for the objective function. Stage 3, the framework determines the optimal grouping of manipulators from the large pool of optimized designs. This process includes selecting the optimal number of groups and allocating the appropriate number of manipulators to each group. The objective is to minimize performance deviation from the original optimal designs and overall manufacturing costs. Given the combinatorial complexity of possible grouping scenarios, a global optimization approach, such as the Non-dominated Sorting Genetic Algorithm II (NSGA-II) navigates the solution space efficiently. The following sections provide a detailed explanation of each stage within the proposed framework.

\newpage

\subsection{Stage 1: Optimal design generation}
\begin{figure*}[b]
\centering
 \includegraphics[width=1\textwidth]{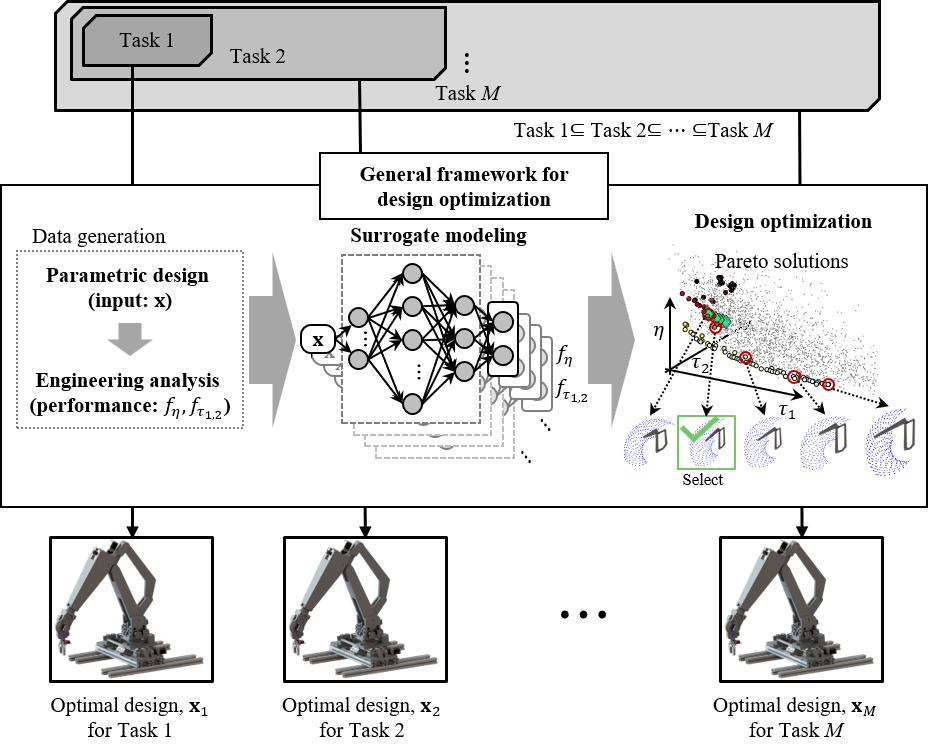}
\caption{Overview of the Stage 1 process for generating optimal manipulator designs tailored to a range of target tasks. Tasks are hierarchically structured based on their workspace requirements, and surrogate modeling is utilized to optimize the designs for given constraints efficiently.}
\label{CH6_fig_stage1}
\vspace{-9pt}
\end{figure*}

In Stage 1, the quasi-serial manipulator optimization framework generates optimal manipulator designs tailored to various tasks~\cite{lee2024multi}. The process of generating optimal manipulator designs using neural network-based surrogate modeling is shown in Fig.~\ref{CH6_fig_stage1}. The methodology comprises four key stages: data generation, engineering analysis, surrogate modeling, and design optimization.

\textbf{1. Data generation:} 
The initial phase involves defining a parametric design space (\textbf{x}), where a set of design variables represents each manipulator. 

\textbf{2. Engineering analysis:} 
Once the design candidates are generated, dynamic analysis is performed to evaluate key engineering performance metrics. Specifically, the workspace coverage efficiency (\(f_{\eta}\)) and the required payload torque (\(f_{\tau_{1,2}}\)) are computed by engineering software. 

\textbf{3. Surrogate modeling:}  
Using the generated dataset, a neural network-based surrogate model is trained to approximate the complex, nonlinear mapping between design parameters and performance outputs.

\textbf{4. Design optimization:}  
With the surrogate model in place, a multi-objective optimization algorithm is employed to identify Pareto-optimal solutions. The objective is to maximize workspace efficiency while minimizing actuation torque requirements. The optimization process yields a diverse set of optimal designs tailored to specific task constraints. By leveraging Pareto filtering, the most suitable design configurations are systematically selected, ensuring they align with functional and manufacturing constraints.

\begin{figure*}[thb]
\centering
 \includegraphics[width=0.5\textwidth]{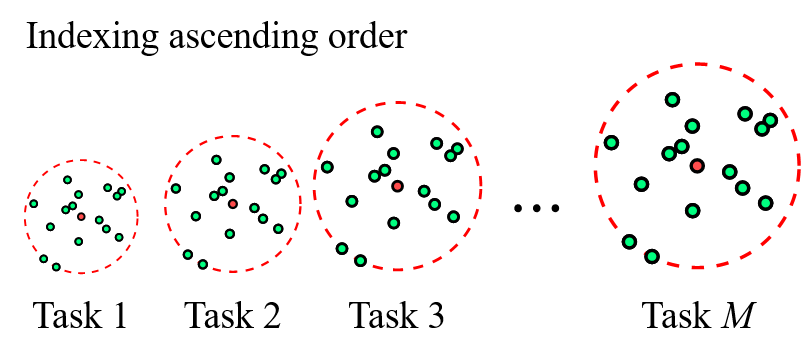}
\caption{Task indexing is input for generating optimal design solutions for $M$ optimal quasi-serial manipulators.}
\label{CH6_fig_stage1_ascending_order}
\vspace{-9pt}
\end{figure*}

\begin{figure*}[thb]
\centering
 \includegraphics[width=0.5\textwidth]{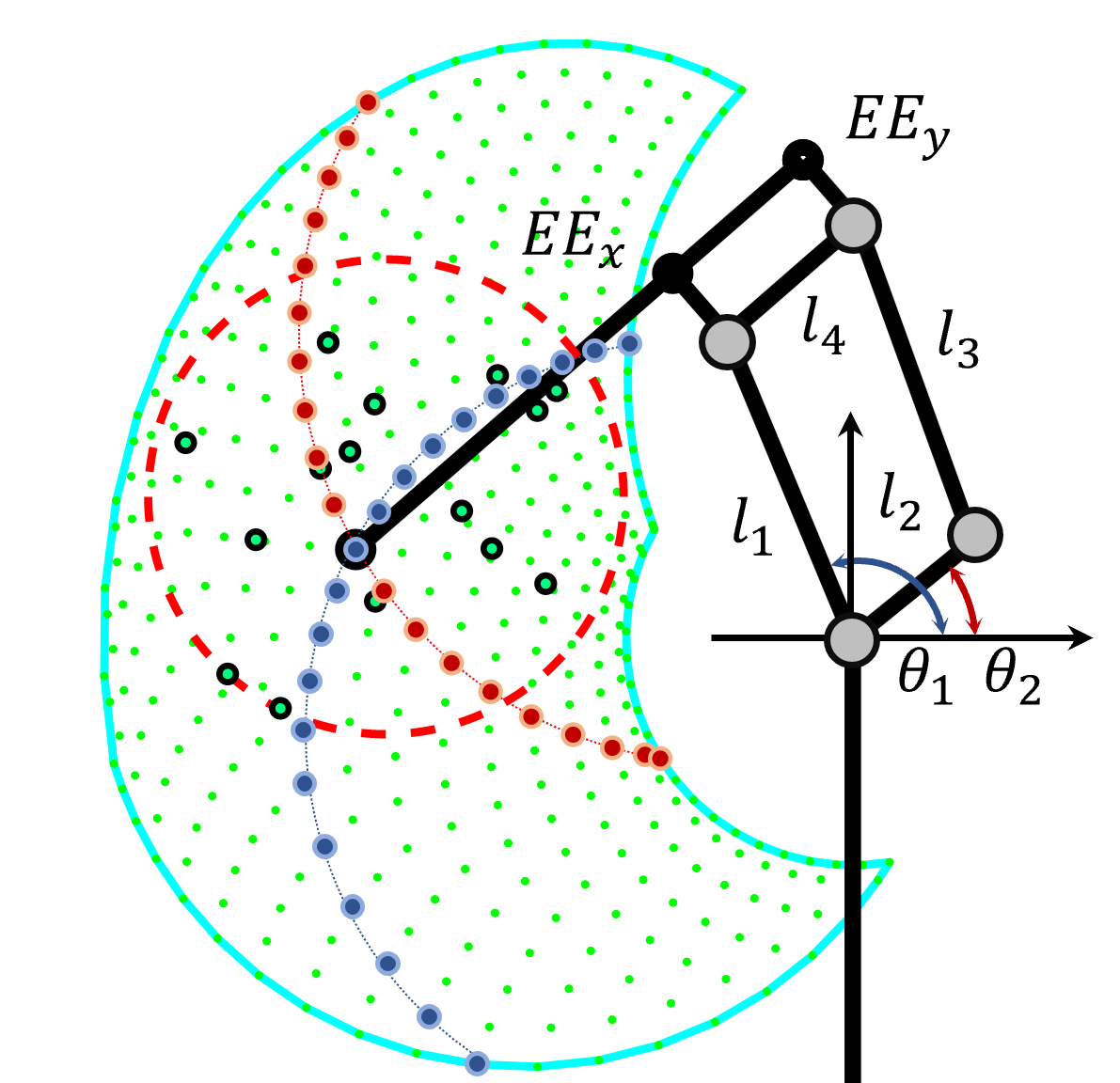}
\caption{Configuration of design variables for the quasi-serial manipulator.}
\label{CH6_fig_stage1_manipulator_DV}
\vspace{-9pt}
\end{figure*}

The assumption for defining a large number of task regions is based on the following two factors, leading to the target task boundary being represented as a circular region:
\begin{enumerate}
    \item To ensure that the manipulator can reach all task points, the workspace is defined as a circle with a radius determined by the farthest task point from the center of the task region(as shown in Fig.~\ref{CH6_fig_stage1_ascending_order}).\\
    
    \item The workspace of a manipulator varies depending on design parameters, installation position, and orientation. Despite these variations, defining the target task region as a circle provides a standardized metric for evaluating the manipulator's ability to perform tasks in all directions. Each manipulator's workspace is optimized to fully encompass the circular target task region, ensuring comprehensive coverage of the designated operating area. The design variables that define the mechanism of the quasi-serial manipulator are illustrated in Fig.~\ref{CH6_fig_stage1_manipulator_DV}.
    
\end{enumerate}

Unlike conventional optimization approaches that focus on a single task, this study aims to address diverse task requirements by efficiently managing a large set of designs and developing a modular design optimization strategy. Stage 1 generates $1,026$ optimal designs for different tasks to achieve this. This number is represented as $M = 1,026$, indicating the total number of manipulators optimized for their respective tasks.

Stage 1 generates optimal designs tailored to diverse tasks, where each task is defined by its workspace and payload requirements. The manipulators are optimized to meet these task conditions, and for simplicity, this study assumes a uniform payload requirement across all tasks. As a result, designs with larger workspaces can perform tasks requiring smaller task regions. The manipulator generation procedure is illustrated in Fig.~\ref{CH6_fig_stage1}. To facilitate systematic grouping, the target tasks are sorted in ascending order based on their workspace size. This sorting ensures that as the workspace increases, the design hierarchy forms a nested structure. Mathematically, this relationship can be expressed as:
\[
\text{Task}~1 \subseteq \text{Task}~2 \subseteq \cdots \subseteq \text{Task}~M.
\]
This hierarchical structure provides a foundation for efficiently leveraging design similarities during grouping. Additionally, it clearly delineates the boundaries of each group, making it easier to identify where grouping should be applied across the models. For any pair of designs \(\textbf{x}_\alpha\) and \(\textbf{x}_\beta\) where \(1 \leq \alpha < \beta \leq M\), the nested structure implies that \(\textbf{x}_\beta\) can cover all tasks that \(\textbf{x}_\alpha\) can perform. In other words, the workspace and payload capabilities of \(\textbf{x}_\beta\) encompass those of \(\textbf{x}_\alpha\). This property is particularly advantageous for modularization, as designs with larger workspaces can serve as representatives for groups containing designs with smaller workspaces, thereby reducing the total number of unique designs required without sacrificing task coverage.

The manipulator design variables are defined as follows:
\[
\textbf{x} = [l_1, l_2, l_3, l_4, EE_x, EE_y, SF],
\]
where \( l_1, l_2, l_3, l_4 \) represent link lengths, \( EE_x, EE_y \) denote the end-effector positions, and \( SF \) indicates the scale factor based on the task region boundary. In Stage 1, these variables are extended to represent large sets of designs. Each variable is defined as:
\begin{align*}
\textbf{x}_j &= [x_{j,1}, x_{j,2}, x_{j,3}, x_{j,4}, x_{j,5}, x_{j,6}, x_{j,7}], \\ 
x_{j,1} &= l_1, \; x_{j,2} = l_2, \; x_{j,3} = l_3, \; x_{j,4} = l_4, \; x_{j,5} = EE_x, \; x_{j,6} = EE_y, \; x_{j,7} = SF.
\end{align*}
where \( j \) represents the index of the optimal designs sorted by target task region size. The required payload torques for each design are defined as \( \tau_{j,1} \) and \( \tau_{j,2} \), which vary across designs. These design variables and torque values serve as objectives for grouping in Stage 3.







\subsection{Stage 2: Cost function definition}
In Stage 2, the cost function is defined to establish a foundation for the optimization process. Manufacturing costs can vary significantly depending on the maturity of production technology and the application environment, making it essential to configure an initial layout tailored to specific manufacturing conditions. When designing large-scale engineering systems, there is significant potential to achieve cost advantages through economies of scale. In a mass production system, the unit manufacturing cost is relatively high at the initial production stage due to fixed setup costs. However, as production volume increases, the cost curve for mass production intersects with that of individual production at a break-even point (\( \text{BEP} \)), beyond which mass production becomes more cost-efficient. In contrast, the unit manufacturing cost for individual production remains almost constant regardless of production volume. For mold-based manufacturing systems, the fixed initial investment cost is spread over larger production volumes, causing the unit manufacturing cost to decrease significantly. Ultimately, mass production systems exhibit unit costs substantially lower than individual production, converging toward the material cost level.

Manufacturing cost curves are highly dynamic and vary depending on the application, material properties, and broader socioeconomic conditions~\cite{wright1936factors, alptekinouglu2008mass, yankovy2020economic}. As a result, it is challenging to apply a universal model that accurately represents all scenarios. Therefore, this study assumes a decreasing exponential function to define the cost curve for ease of use within the proposed framework. However, the cost function remains adjustable, allowing the cost curve to take different forms to suit various applications and manufacturing scenarios. To quantitatively model these economies of scale and evaluate the trade-offs in different production scenarios, this study defines the manufacturing cost curve using the following parameters:

\begin{itemize}
    \item $\omega_0$: Initial manufacturing cost, including the setup costs for mass production facilities and associated overhead. This involves fixed infrastructure costs such as tooling, presses, and molds. \\  
    \item $\omega_{\min}$: Minimum manufacturing cost, representing the lower bound of per-unit production costs as production scales up. This includes material costs and the minimum operational expenses required for production. \\  
    \item $\text{BEP}$: Break-even point, where the per-unit cost of mass production equals that of individual production. Beyond this point, producing in larger quantities becomes more cost-effective, with costs converging toward $\omega_{\min}$ as production scales up. \\  
\end{itemize}

Traditionally, manufacturing cost curves are defined based on absolute per-unit production costs~\cite{jung2023additive}. However, this study adopts a relative cost ratio to compare the cost reduction benefits of mass production with individual production. By normalizing the cost of individual production as 1, the relative cost reduction achieved through mass production can be directly evaluated. Consequently, the cost function is determined by the parameters $\omega_0$, $\omega_{\min}$, and $\text{BEP}$, allowing it to adapt to varying levels of technological maturity and manufacturing environments. This study utilizes an exponential cost function incorporating the damping coefficient $\kappa$, defined as follows:
\begin{equation}\label{CH6_cost_function}
\omega(n) =
\begin{cases}
1 & (n \leq \text{BEP}), \\
\omega_0 e^{-\kappa n} + \omega_{\min} & (n > \text{BEP}).
\end{cases}
\end{equation}

Where $n$ represents the number of units produced. It is used to evaluate at what production volume mass production becomes more cost-effective than individual production relative to the break-even point. In Stage 3, $n$ corresponds to the number of manipulators assigned to each group, which serves as an input for cost optimization within the grouping strategy. Additionally, $\kappa$ is the damping coefficient, calculated from $\omega_0$, $\omega_{\min}$, and $\text{BEP}$ using the following derivation:
\begin{align} \label{CH6_eq_kappa}
\omega_0 e^{-\kappa \cdot \text{BEP}} + \omega_{\min} = 1, \\
\omega_0 e^{-\kappa \cdot \text{BEP}} = 1 - \omega_{\min}, \notag\\
-\kappa \cdot \text{BEP} = \ln \left( \frac{1 - \omega_{\min}}{\omega_0} \right), \notag\\
\therefore \kappa = \frac{1}{\text{BEP}} \ln \left( \frac{\omega_0}{1 - \omega_{\min}} \right). \notag
\end{align}

\begin{figure*}[h!]
\centering
 \includegraphics[width=0.5\textwidth]{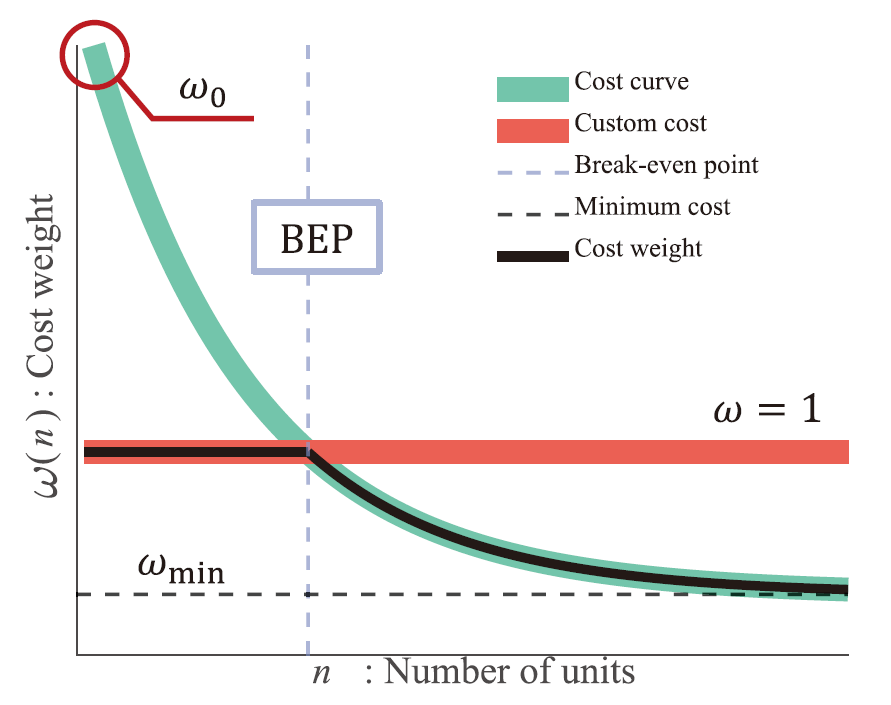}
\caption{Cost function curve illustrating the transition from custom production to mass production.}
\label{CH6_fig_cost_function}
\vspace{-9pt}
\end{figure*}

Using the above Eq.~\eqref{CH6_cost_function} and \eqref{CH6_eq_kappa}, $\kappa$ can be calculated, serving as a key component in defining the cost function for the optimization process. This cost function, illustrated in Fig.~\ref{CH6_fig_cost_function}, is utilized in Stage 3 to evaluate grouped designs by capturing the trade-offs between cost and performance during optimization.

\newpage
\subsection{Stage 3: Modular design optimization}
\begin{figure*}[b!]
\centering
 \includegraphics[width=1\textwidth]{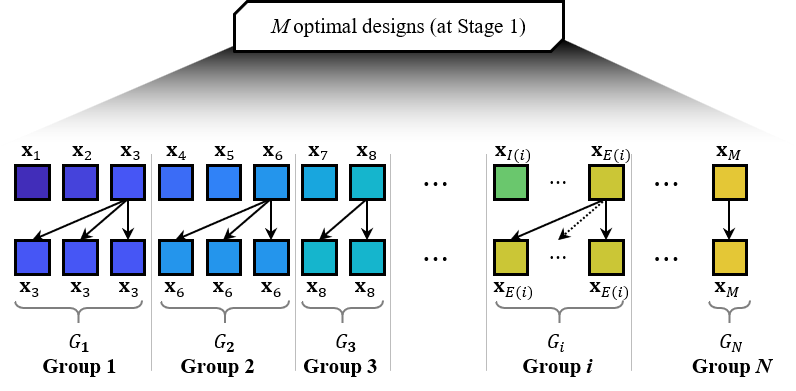}
\caption{Grouping of \( M \) optimal manipulator designs into \( G_N \) groups. The figure illustrates the process of assigning designs to groups and the subsequent unification strategy within each group, where designs from \( I(i) \) to \( E(i) \) are consolidated into a standardized design \( \mathbf{x}_{E(i)} \).}
\label{CH6_fig_stage3_1}
\vspace{-9pt}
\end{figure*}

Stage 3 leverages the optimal designs generated in Stage 1 and the cost function defined in Stage 2 to implement a modular design strategy. This involves partitioning the \( M \) optimal manipulator designs into \( N \) groups, where each group minimizes manufacturing costs while maintaining minimal performance deviation from the original designs. To explain the proposed strategy clearly, Fig.~\ref{CH6_fig_stage3_1} introduces the notation used for grouping. In Stage 1, a total of \( M \) optimal manipulator designs were generated. These designs are divided into \( N \) groups, and the representative design for each group is determined based on the manipulator with the largest target task region within the group. The indices of manipulators within the \( i \)-th group, denoted as \( G_i \), range from \( I(i) \) to \( E(i) \).

The starting index \( I(i) \) and the ending index \( E(i) \) for each group are defined as follows:
\begin{align}
I(i) &= 
\begin{cases} 
1, & (i = 1), \\
1 + \displaystyle\sum_{q=1}^{i-1} G_q, & (i > 1),
\end{cases} \label{CH6_initial_index} \\
E(i) &= I(i) + G_i - 1. \label{CH6_final_index}
\end{align}

Where, \( I(i) \) represents the starting index of the \( i \)-th group (in Eq.~\eqref{CH6_initial_index}), and \( E(i) \) represents the ending index(in Eq.~\eqref{CH6_final_index}), determined by adding the size of the group \( G_i \) to \( I(i) \). The representative design variable for each group, which is used for grouping, is selected as \( \mathbf{x}_{E(i)} \), corresponding to the manipulator with the largest target task region in the group. Figure~\ref{CH6_fig_stage3_1} illustrates the proposed grouping process, where designs within a range from \( I(i) \) to \( E(i) \) are replaced by the standardized design \( \mathbf{x}_{E(i)} \).

The notations used in this strategy are explained as follows:
\begin{itemize}
    \item \( M \): Total number of optimal manipulator designs generated in Stage 1.
    \item \( N \): Total number of groups into which the designs are partitioned.
    \item \( G_i \): Size (number of manipulators) of the \( i \)-th group.
    \item \( I(i) \): Starting index of the \( i \)-th group.
    \item \( E(i) \): Ending index of the \( i \)-th group.
    \item \( \mathbf{x}_{E(i)} \): Design variable of the representative manipulator for the \( i \)-th group, determined as the manipulator with the largest target task region in the group.
\end{itemize}

\newpage
\subsubsection{Modular design: concept and challenges}
The modular design approach groups individual products into unified designs to streamline manufacturing, reduce costs, and simplify maintenance. While grouping offers advantages such as cost savings through mass production and easier maintenance, it introduces unavoidable performance deviations due to differences between the standardized and individually optimized designs. Thus, optimizing the objective functions for modular design requires addressing the following essential questions:
\begin{itemize}
    \item How many groups should be created to accommodate the required designs? 
    \item How many designs should be assigned to each group?
\end{itemize}

However, this approach also faces significant computational challenges due to the exponential growth in the number of possible groupings as \( M \) and \( N \) increase. Intuitively, determining the number of groups and the number of designs assigned to each group can be considered a combinatorial problem. The total number of ways to partition \( M \) designs into \( N \) groups, where each group contains at least one design, is given by:
\begin{equation} \label{CH6_eq_comb}
\text{Comb}(M, N) = \binom{M-1}{N-1} = \frac{(M-1)!}{(N-1)!(M-N)!}.
\end{equation}
Where, \( \binom{M-1}{N-1} \) represents the number of ways to place \( N-1 \) dividers among \( M-1 \) designs, ensuring that each group contains at least one design and that the total number of designs across all groups equals \( M \).

As \( M \) and \( N \) increase, the number of possible groupings grows exponentially, making exhaustive evaluation computationally impractical. Equation~\eqref{CH6_eq_comb} demonstrates that the total number of ways to partition \( M \) designs into \( N \) groups is determined by combinatorial complexity. For instance, when \( M = 20 \), the number of possible groupings reaches 524,288, showing the challenge posed even for relatively small-scale problems. Considering that large-scale engineering design often involves hundreds or thousands of designs, it becomes evident that simultaneously optimizing the number of groups and the designs assigned to each group is computationally infeasible~\cite{wu2020knowledge}. While grouping aims to achieve uniformity among designs, minimizing performance losses while maximizing cost benefits, exploring the entire design space becomes unnecessary as the complexity and diminishing economic returns outweigh potential advantages. Beyond a certain point, excessive grouping configurations fail to deliver meaningful cost benefits, making such approaches inefficient.

\newpage
\subsubsection{Objective functions and constraints}
To achieve economic advantages via economies of scale, beginning with a small number of groups and gradually increasing the group count is computationally efficient. Observing the cost function's behavior as the group count increases helps identify configurations that balance cost efficiency and performance. Instead of simultaneously optimizing all possible group configurations and design assignments, a stepwise exploration strategy is adopted to maximize economies of scale. Starting with a small number of groups, the search incrementally increases \( N \), enabling efficient solution space exploration while managing computational complexity. This approach systematically refines grouping configurations, prioritizing cost savings and performance balance.
To address the computational complexity of grouping \( M \) designs into \( N \) groups, the multi-objective optimization algorithm NSGA-II is employed. The optimization problem is formulated to minimize objectives: manufacturing cost and payload torque deviation, as shown in Fig.~\ref{CH6_fig_stage3_2}. The first objective, manufacturing cost (\( C \)), is assumed to be proportional to the material usage, which is directly related to the size of each manipulator's components. The size of the manipulator in each group is defined as:

\begin{figure*}[h!]
\centering
 \includegraphics[width=1\textwidth]{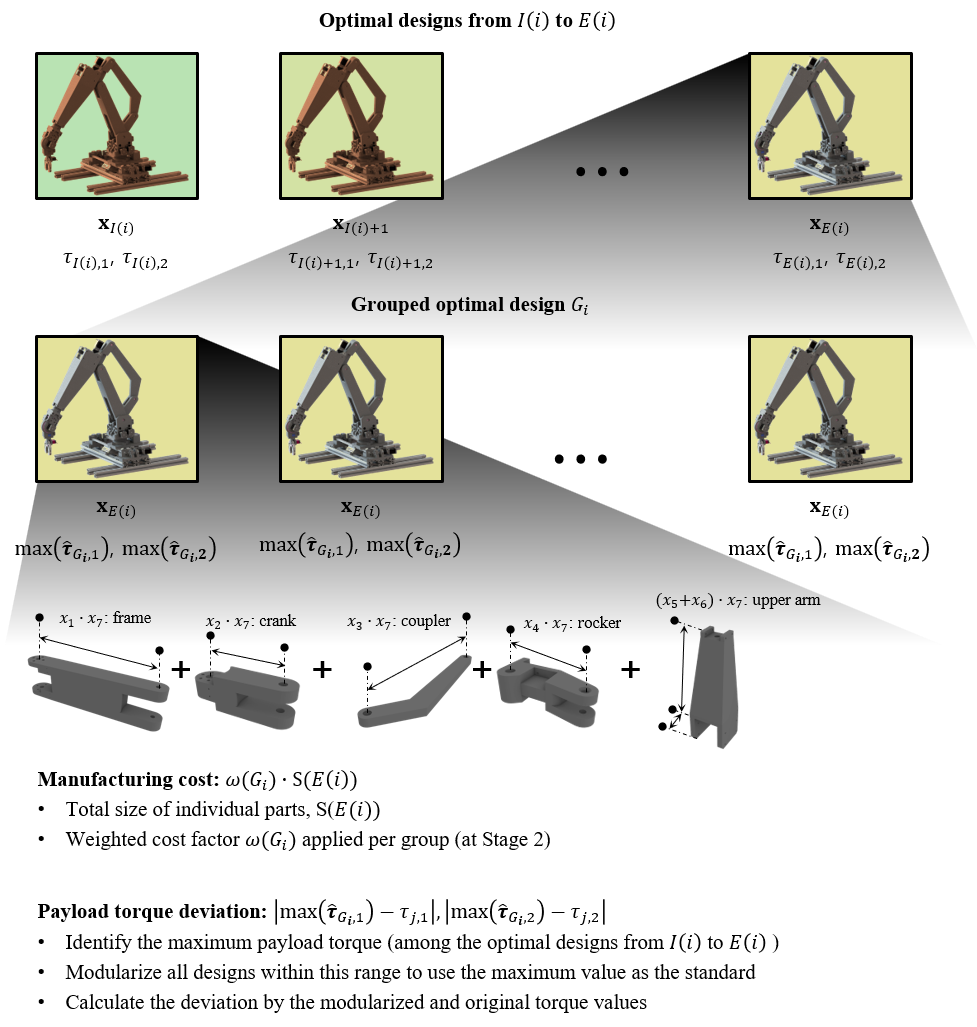}
\caption{Grouping process for optimal manipulator designs. The figure shows the transformation of optimal designs from \( I(i) \) to \( E(i) \) into a single grouped design \( \mathbf{x}_{E(i)} \) within each group \( G_i \). The manufacturing cost is computed using \( \omega(G_i) \cdot S(E(i)) \), where \( S(E(i)) \) represents the total size of individual parts. The payload torque deviation is evaluated as \( |\text{max}(\bm{\hat{\tau}}_{G_i,1}) - \tau_{j,1}| \) and \( |\text{max}(\bm{\hat{\tau}}_{G_i,2}) - \tau_{j,2}| \) to quantify the difference between standardized and original torque values.}
\label{CH6_fig_stage3_2}
\vspace{-9pt}
\end{figure*}

\begin{align}
S(E(i)) &= \sum_{k=1}^{6} (x_{E(i), k} \cdot x_{E(i), 7}). \label{CH6_size}
\end{align}
In Eq.~\eqref{CH6_size}, \( x_{E(i), k} \) represents the \( k \)-th design variable of the manipulator chosen as the representative for the \( i \)-th group, while \( x_{E(i), 7} \) denotes the scale factor.

Using the cost function curve defined in Stage 2 (in Fig.~\ref{CH6_fig_cost_function}), the weighted cost for each group is computed based on the number of manipulators in the group (\( G_i \)). The total manufacturing cost across all groups is then defined as:
\begin{align}
C(\textbf{G}_{N}) &= \sum_{i=1}^{N} \sum_{j=I(i)}^{E(i)} \big( \omega(G_i) \cdot S(E(i)) \big), \label{CH6_obj_1}
\end{align}
where \( \omega(G_i) \) is the weight factor applied based on the group size.

The second and third objectives, payload torque deviation (\(\Delta\Gamma_{1},\Delta\Gamma_{2}\)), arises from the grouping process. For each group, the maximum payload torque required among all designs within the group is selected as the representative value:
\begin{align}
\bm{\hat{\tau}}_{G_i, 1} &= \{\tau_{j,1} \;|\; j \in [I(i), E(i)]\}, \label{CH6_max_tau_1} \\  
\bm{\hat{\tau}}_{G_i, 2} &= \{\tau_{j,2} \;|\; j \in [I(i), E(i)]\}. \label{CH6_max_tau_2}
\end{align}
The deviations between the selected maximum payload torque and the original torques of the individual designs are minimized.  
For the \( i \)-th group, the payload torque is defined in Eqs.~\eqref{CH6_max_tau_1}, and ~\eqref{CH6_max_tau_2}, and the maximum value among them is selected. Thus, the total torque deviations across all groups are expressed as:
\begin{align}
\Delta\Gamma_{1}(\textbf{G}_{N}) &= \sum_{i=1}^{N} \sum_{j=I(i)}^{E(i)} \left| \max(\bm{\hat{\tau}}_{G_i, 1}) - \tau_{j,1} \right|,\label{CH6_obj_2} \\
\Delta\Gamma_{2}(\textbf{G}_{N}) &= \sum_{i=1}^{N} \sum_{j=I(i)}^{E(i)} \left| \max(\bm{\hat{\tau}}_{G_i, 2}) - \tau_{j,2} \right|. \label{CH6_obj_3}
\end{align}

\begin{figure*}
\centering
 \includegraphics[width=0.5\textwidth]{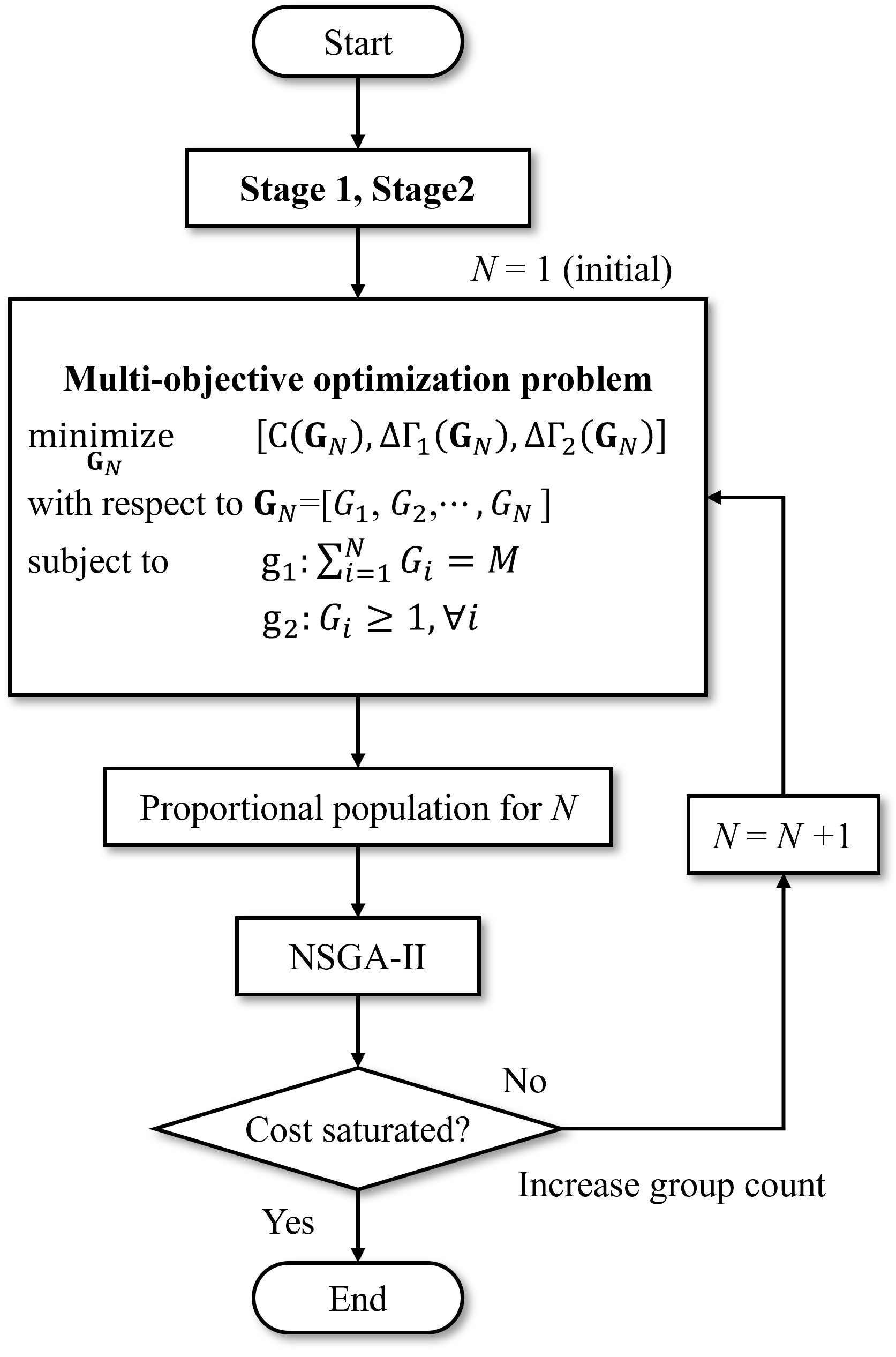}
\caption{Step-by-step procedure for modular design layout optimization with incremental grouping strategy until cost saturation}
\label{CH6_fig_MOO}
\vspace{-9pt}
\end{figure*}

To efficiently explore the solution space, the number of groups (\( N \)) is incrementally increased, starting from \( N = 1 \), with the NSGA-II algorithm applied iteratively for each \( N \) (as shown in Fig.~\ref{CH6_fig_MOO}). This process accumulates Pareto-optimal solutions, enabling an evaluation of the trade-offs between cost and performance. The concept of the break-even point (\( \text{BEP} \)) defined in Stage 2 serves as a critical guideline for the grouping strategy, as dividing the total number of designs \( M \) by \( \text{BEP} \) provides an estimate of the average number of manipulators that should be assigned to each group to achieve cost benefits. If the number of manipulators in a group falls below this threshold, the cost advantage diminishes, potentially favoring individual production. Furthermore, as the number of groups increases beyond the threshold implied by the \( \text{BEP} \), the cost function saturates, indicating that further grouping provides little to no improvement in the optimization objectives and may not be economically meaningful.

Thus, the optimization problem in Stage 3 can be formulated as a multi-objective optimization (MOO) problem, incorporating the objective functions given in Eqs.~\eqref{CH6_obj_1}, \eqref{CH6_obj_2}, and \eqref{CH6_obj_3}, along with the associated constraints. It can be summarized as follows:

\begin{equation}
\setlength{\arraycolsep}{0pt} 
\begin{array}{ll} \label{CH6_eq_obj_modular_design}
\underset{\textbf{G}_{N}}{\text{minimize}} &\quad [C(\textbf{G}_{N}),\Delta\Gamma_{1}(\textbf{G}_{N}),\Delta\Gamma_{2}(\textbf{G}_{N})] \\

\text{with respect to} 
&\quad \textbf{G}_{N} = [G_1, G_2, \cdots, G_{N}] \\

\text{subject to} 
&\quad g_1: \displaystyle\sum_{i=1}^{N} G_i = M \\
&\quad g_2: G_i \geq 1, \; \forall i\\

\text{where} 
&\quad C(\textbf{G}_{N}) = \displaystyle\sum_{i=1}^{N} \sum_{j=I(i)}^{E(i)}(\omega(G_i) \cdot S(E(i))) \\
&\quad \Delta\Gamma_{1}(\textbf{G}_{N}) = \displaystyle\sum_{i=1}^{N} \sum_{j=I(i)}^{E(i)}|\text{max}(\bm{\hat{\tau}}_{G_i, 1}) - \tau_{j,1}| \\
&\quad \Delta\Gamma_{2}(\textbf{G}_{N}) = \displaystyle\sum_{i=1}^{N} \sum_{j=I(i)}^{E(i)}|\text{max}(\bm{\hat{\tau}}_{G_i, 2}) - \tau_{j,2}| \\

&\quad I(i) = 
    \begin{cases} 
        1 & (i = 1) \\
        1+\displaystyle\sum_{q=1}^{i-1} G_q & (i > 1)
    \end{cases} \\
&\quad E(i) = I(i) + G_{i} - 1\\
\\
&\quad  \omega(G_i) = 
    \begin{cases}
        1 & (G_i \leq \text{BEP}) \\
        \omega_0 e^{-\kappa G_i} + \omega_{\min} & (G_i > \text{BEP})
    \end{cases}\\
\\
&\quad  S(E(i)) = \displaystyle\sum_{k=1}^{6} (x_{E(i), k} \cdot x_{E(i), 7})\\
&\quad  \bm{\hat{\tau}}_{G_i, 1} = \{\tau_{j,1} \;|\; j \in [I(i), E(i)]\} \\
&\quad  \bm{\hat{\tau}}_{G_i, 2} = \{\tau_{j,2} \;|\; j \in [I(i), E(i)]\}\\

\end{array}
\end{equation}
\noindent Constraints (\( g_1, g_2 \)): Guarantee all manipulators are assigned to groups while maintaining non-empty group sizes.

\section{Results and discussion}\label{sec4}
In large-scale engineering design scenarios, the outcomes are influenced by the objective functions defined by designers and the parameters of the considered cost curve. These factors can be dynamically adjusted depending on the timing of the design requirements and the specific conditions of the application. This section analyzes the impact of varying key parameters of the cost curve proposed in Stage 2 on the optimization results, using a large-scale design scenario for industrial manipulators as a case study. This analysis evaluates the proposed modular design framework to balance design costs and performance trade-offs, demonstrating its adaptability and practicality in large-scale design problems.


In mass production systems, the initial manufacturing cost (\( \omega_0 \)), the minimum cost converging during mass production (\( \omega_{\min} \)), and the break-even point (\( \text{BEP} \)) serve as key variables in manufacturing strategy. The \( \text{BEP} \) represents the production volume at which mass production becomes more economically advantageous than individual production, which can vary significantly depending on the manufacturing environment and applied technologies. In particular, determining \( \text{BEP} \) is highly dependent on the maturity of manufacturing technologies and the characteristics of the application, often exhibiting substantial deviations, making its precise estimation a challenging task. To analyze the influence of \( \omega_0 \), \( \omega_{\min} \), and \( \text{BEP} \) on the optimization outcomes, a case study was conducted under a simplified manufacturing environment with several predefined conditions.

To objectively compare the performance of the proposed application, \( C_{\text{origin}}, \Gamma_{\text{origin},1}, \Gamma_{\text{origin},2} \) are introduced. These values represent the total incurred cost and the sum of the required payload torques when each of the \( M \) design candidates generated in Stage 1 is manufactured independently. The respective values are computed using the following equations:

\begin{align}\label{CH6_eq_origin}
     C_{\text{origin}} &= \sum_{i=1}^{M} S(i), \quad 
\Gamma_{\text{origin,1}} = \sum_{i=1}^{M} \tau_{i,1}, \quad
\Gamma_{\text{origin,2}} = \sum_{i=1}^{M} \tau_{i,2}.
\end{align}

Subsequently, for each case, the objective function values obtained by applying the modular design framework are normalized using the initial cost and required payload torque values computed in Eq.~\eqref{CH6_eq_origin}. To facilitate this, \( C_{\text{ratio}}, \Gamma_{\text{ratio},1}, \Gamma_{\text{ratio},2} \) are defined, allowing for an objective comparison of the optimization results against the initial design candidates. These values are formulated as follows:

\begin{align}\label{CH6_eq_origin2}
     C_{\text{ratio},N}  &= \frac{C(\textbf{G}^*_{N})}{C_{\text{origin}}}, \quad 
\Gamma_{\text{ratio,1},N} = \frac{\Delta\Gamma_{1}(\textbf{G}^*_{N})}{\Gamma_{\text{origin,1}}}, \quad
\Gamma_{\text{ratio,2},N} = \frac{\Delta\Gamma_{2}(\textbf{G}^*_{N})}{\Gamma_{\text{origin,2}}}.
\end{align}

Where $\textbf{G}^*_N$ represents the optimal design assignment for each group computed via Eq.~\eqref{CH6_eq_obj_modular_design} optimization process as the number of groups (\( N \)) increases incrementally by one. The number of groups (\( N \)) reaches a maximum based on the total number of design candidates (\( M \)), and an exhaustive search across all possible combinations would require consideration up to \( N = M \). However, searching all combinations is computationally and financially impractical when aiming for cost efficiency through grouping. The break-even point (BEP) for each case is utilized to address this. The BEP denotes the production quantity at which mass production becomes more economical than individual production. The average number of designs per group can be estimated by dividing the total number of design candidates (\( M \)) by the BEP. This shows that, beyond a certain threshold, economies of scale can lead to cost reduction.

Conversely, suppose excessive group segmentation (i.e., a sufficiently high \( N \)) is required. In that case, the number of designs assigned to each group may fall below the threshold necessary for the predefined cost curve (in Eq.~\eqref{CH6_cost_function}) to be applicable, resulting in individual production costs being applied instead. This scenario can lead to a saturation state, where the cost becomes comparable to or even higher than producing all designs individually at a sufficiently high \( N \). This saturation state is defined as Sat and is expressed as:


\begin{align}\label{CH6_eq_sat}
     \text{Sat} &= \frac{M}{\text{BEP}}.
\end{align}

To ensure computational efficiency, the Sat value is used as a criterion for estimating the appropriate number of groups. The number of NSGA-II iterations (\( N \)) is set to the rounded integer value twice the Sat value. In this process, the optimal grouping is evaluated by verifying whether \( C_{\text{ratio}} \) converges to 1 based on the Sat threshold.

\newpage
\subsection{Impact of the break-even point (\( \text{BEP} \)) on objective functions}
\begin{figure*}[h!]
\centering
 \includegraphics[width=1\textwidth]{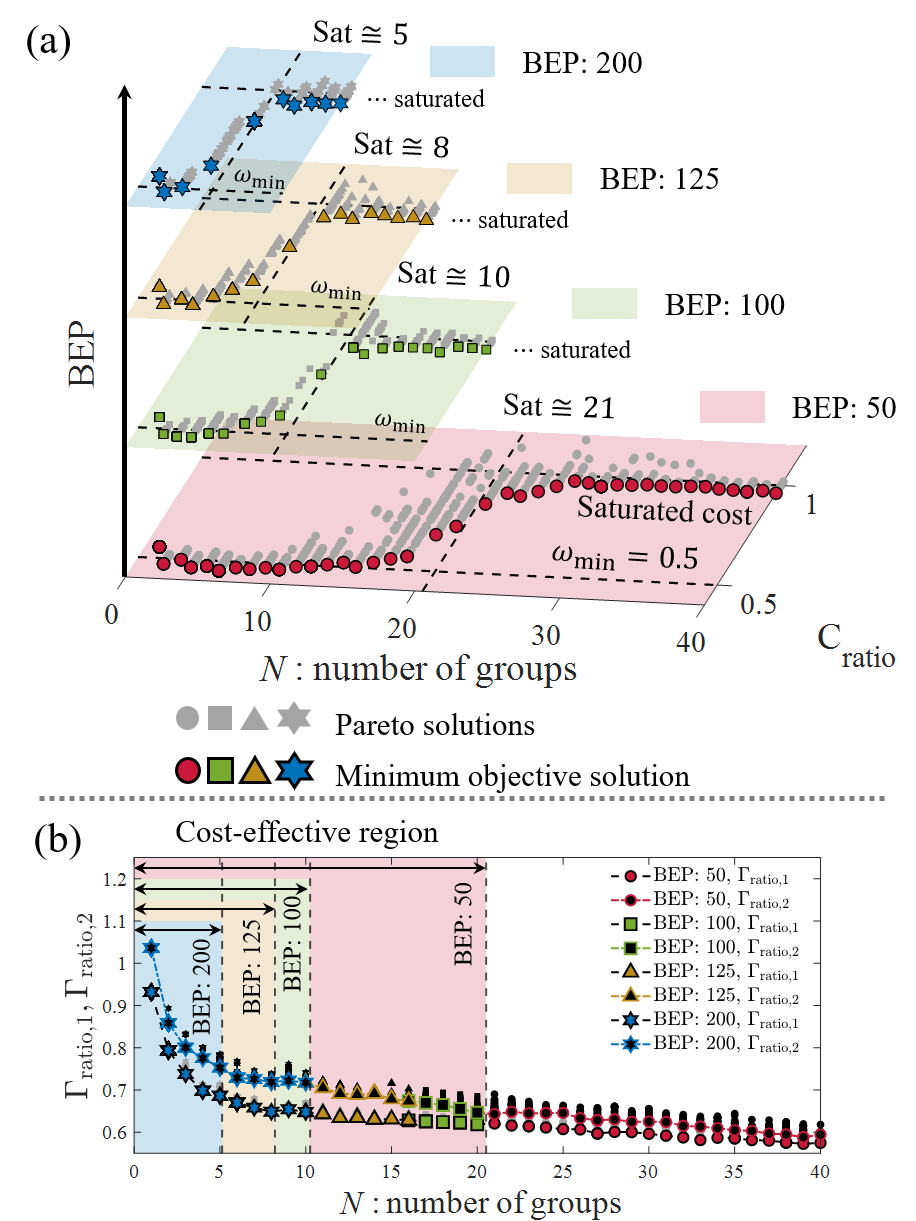}
\caption{(a) Displays the cost curve under varying \( \text{BEP} \) values, showing the saturation where \( C_{\text{ratio}} \) converges to 1 for \( N \) greater than the Sat value. (b) Displays the \( \Gamma_{\text{ratio,1}} \) and \( \Gamma_{\text{ratio,2}} \), indicating the payload torque deviation decreases as  \( N \) increases.}
\label{CH6_fig_result1}
\vspace{-9pt}
\end{figure*}

To analyze the effect of variations in the break-even point (\( \text{BEP} \)) on the number of groups (\( N \)) and manufacturing costs, \( \text{BEP} \) was gradually increased while keeping the initial manufacturing cost (\( \omega_0 \)) and the minimum cost (\( \omega_{\min} \)) fixed. The detailed settings for this analysis are provided in Table~\ref{CH6_table_case_study}. The results indicate that as \( \text{BEP} \) increases, the number of groups (\( N \)) required for \( C_{\text{ratio}} \) to converge to 1 tends to decrease. This trend suggests that a higher \( \text{BEP} \) reduces the average number of groups needed to achieve economies of scale, as illustrated in Fig.~\ref{CH6_fig_result1} (a). When \( C_{\text{ratio}} \) is less than 1, a cost advantage exists compared to producing each design independently. However, as \( \text{BEP} \) increases, the potential for cost reduction diminishes, implying that at high \( \text{BEP} \) values, the effectiveness of achieving cost efficiency through appropriate grouping may be reduced. Therefore, the economically viable range of group numbers under each condition can be defined as the region where \( N \) remains below the Sat value, which is highlighted in Fig.~\ref{CH6_fig_result1} (b).


In contrast to Fig.~\ref{CH6_fig_result1} (a), Fig.~\ref{CH6_fig_result1} (b) exhibits a clear decreasing trend in \( \Gamma_{\text{ratio,1}} \) and \( \Gamma_{\text{ratio,2}} \) as \( N \) increases. This trend can be interpreted as reducing payload torque deviation as the optimal designs are grouped with finer resolution. Additionally, since the cost curve does not apply to \( \Gamma_{\text{ratio,1}} \) and \( \Gamma_{\text{ratio,2}} \) per group, the values calculated at the same group number under the conditions in Table~\ref{CH6_table_case_study} appear to be highly similar. A comprehensive analysis of these results suggests that the optimal group configuration can be determined based on the inflection point where the slope of \( N \) changes abruptly.

\begin{table}[h]
\centering
\caption{The cost curve parameters for each case are summarized and used in Fig.~\ref{CH6_fig_result1}.}
\label{CH6_table_case_study}
\begin{tabularx}{0.5\linewidth}{c*{5}{>{\centering\arraybackslash}X}} 
\toprule
\Xhline{2pt}
  & (a-1) & (a-2) &  (a-3) & (a-4)  \\
\midrule
$\omega_{0}$   & 10  & 10    & 10  & 10  \\ 
$\omega_{\text{min}}$ & 0.5 & 0.5  & 0.5 & 0.5 \\
    BEP        & 50  & 100   & 125  & 200 \\

\Xhline{2pt}
\bottomrule
\end{tabularx}
\end{table}

\newpage
\newpage

\subsection{Impact of the initial manufacturing cost(\( \omega_0 \)) on objective functions}
\begin{figure*}[h!]
\centering
 \includegraphics[width=1\textwidth]{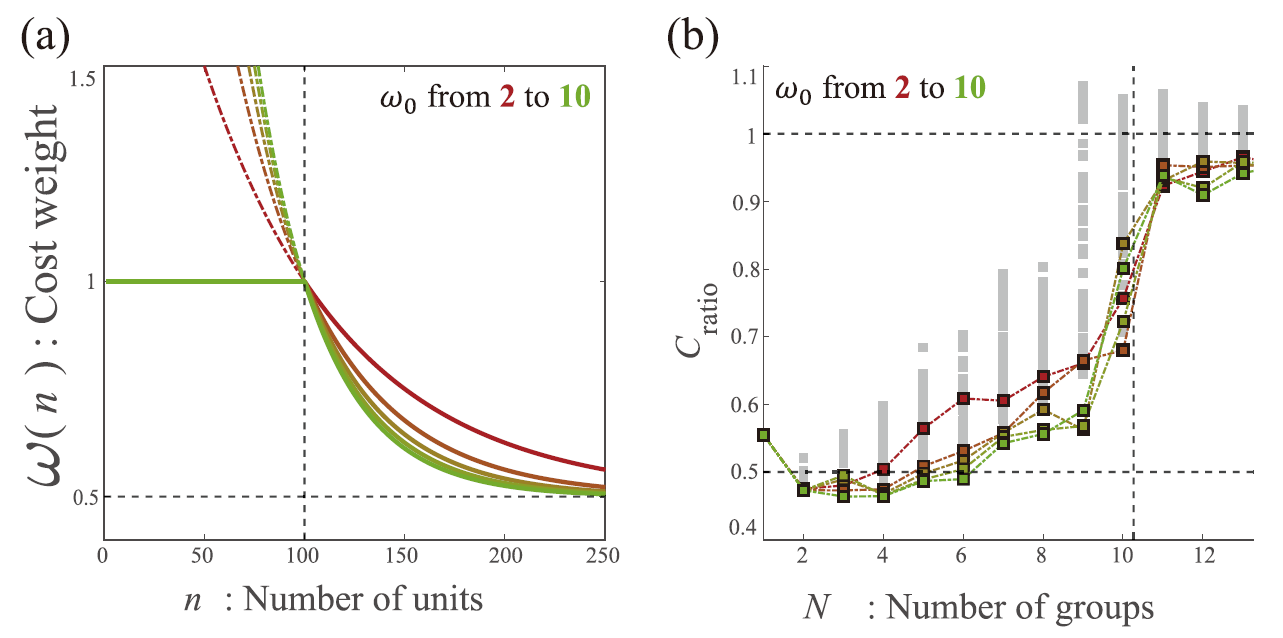}
\caption{Varying \( \omega_0 \) on the cost curve. (a) Illustrates the cost weight \( \omega_0 \) increases, the cost weight decreases rapidly beyond the break-even point. (b) Displays the relationship between \( C_{\text{ratio}} \) and the number of groups (\( N \)), showing that higher \( \omega_0 \) values lead to lower \( C_{\text{ratio}} \) for \( N < \text{Sat} \) and sharper increases in \( C_{\text{ratio}} \) near the Sat line.}
\label{CH6_fig_result2}
\vspace{-9pt}
\end{figure*}

In the previous analysis, the impact of variations in the break-even point (\( \text{BEP} \)) on the objective functions was examined based on the conditions outlined in Table~\ref{CH6_table_case_study}. Among the results presented in Fig.~\ref{CH6_fig_result1}(a-2), the most pronounced change was observed near the Sat value under the condition of \( \text{BEP} = 100 \). Based on this observation, the effect of variations in the initial manufacturing cost (\( \omega_0 \)) on the objective functions and cost curve is analyzed using the conditions of Table~\ref{CH6_table_case_study}(a-2) as a reference. The optimization was conducted while progressively increasing \( \omega_0 \) from 2 to 10, and the results are illustrated in Fig.~\ref{CH6_fig_result2} (a) and (b).


Figure~\ref{CH6_fig_result2} (a) presents the effect of increasing \( \omega_0 \) on the shape of the cost curve. The analysis indicates that as \( \omega_0 \) increases, the cost weight beyond the break-even point (\( n \)) decreases more sharply. This denotes that a higher \( \omega_0 \) emphasizes cost efficiency beyond the break-even point in the grouping strategy.

Figure~\ref{CH6_fig_result2} (b) presents the impact of \( \omega_0 \) variations on the relationship between \( C_{\text{ratio}} \) and the number of groups (\( N \)). As \( \omega_0 \) increases, changes in the cost curve exert a greater influence on \( C_{\text{ratio}} \) near Sat, resulting in a steeper increase. This finding implies that when manufacturing costs drop sharply beyond the break-even point, there is greater potential for cost savings. However, this also indicates that if the number of groups near Sat is not carefully selected in the grouping strategy, the efficiency may decline due to a sudden rise in costs.

\newpage
\subsection{Impact of the minimum manufacturing cost(\( \omega_{\min} \)) on objective functions}

\begin{figure*}[h!]
\centering
 \includegraphics[width=1\textwidth]{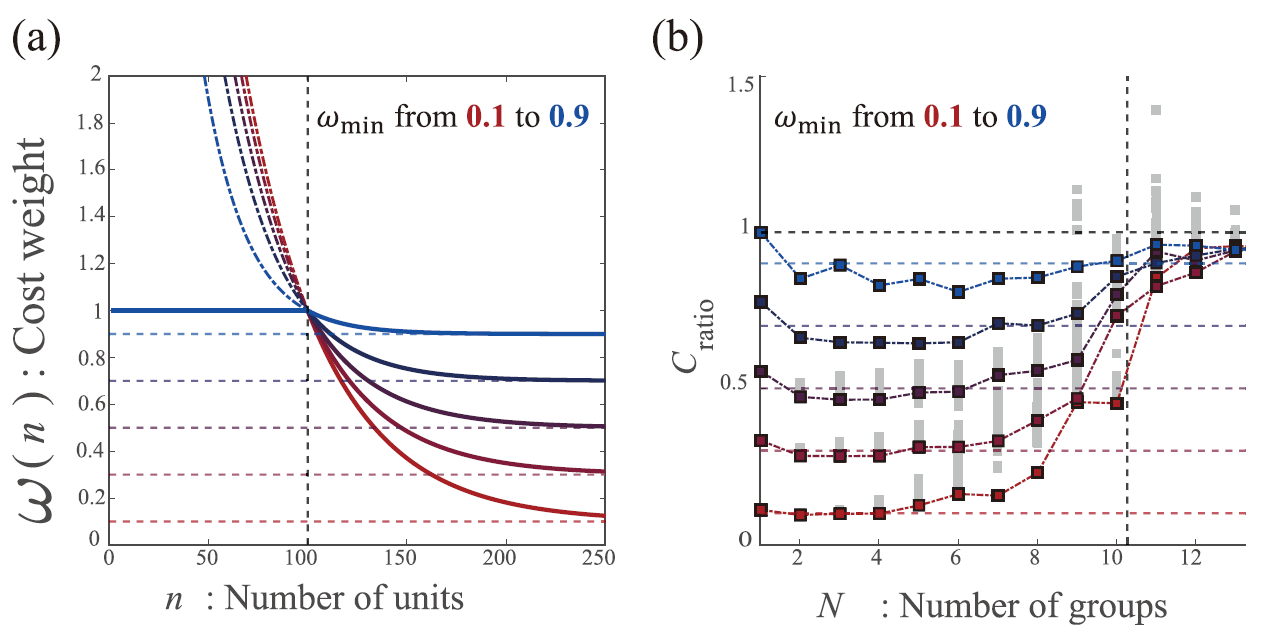}
\caption{Varying \( \omega_{\min} \) on the cost curve. (a) Illustrates the cost weight \( \omega(n) \) as \( \omega_{\min} \) increases, showing a rise in the cost curve beyond the break-even point and an increase in the required \( n \) for sufficient convergence. (b) Displays the relationship between \( C_{\text{ratio}} \) and the number of groups (\( N \)), indicating that higher \( \omega_{\min} \) values lead to increased costs before reaching the Sat line.}
\label{CH6_fig_result3}
\vspace{-9pt}
\end{figure*}

To analyze the impact of variations in the minimum manufacturing cost (\( \omega_{\min} \)) on the cost curve and objective functions, optimization was performed while gradually increasing \( \omega_{\min} \) from 0.1 to 0.9 under the conditions of Table~\ref{CH6_table_case_study} (a-2). The results are presented in Fig.~\ref{CH6_fig_result3} (a) and (b).

Figure~\ref{CH6_fig_result3} (a) presents the effect of increasing \( \omega_{\min} \) on the shape of the cost curve. The analysis indicates that as \( \omega_{\min} \) increases, the cost weight beyond the break-even point (\( n \)) gradually increases. This indicates that even after the cost stabilizes in the mass production phase, a higher \( \omega_{\min} \) maintains an elevated cost level.

Figure~\ref{CH6_fig_result3} (b) presents the effect of \( \omega_{\min} \) variations on \( C_{\text{ratio}} \) as a function of the number of groups (\( N \)). Notably, when \( \omega_{\min} \) exceeds 0.5, its impact on \( C_{\text{ratio}} \) becomes more pronounced. In this case, the Pareto solutions were constrained by conditions necessary to achieve sufficient objective function improvements. Unlike other conditions, Pareto-optimal samples identified within this range (\(\omega_{\min}>0.5\)) were concentrated near the minimum value, limiting the exploration of diverse solutions. The results indicate that lower \( \omega_{\min} \) values lead to relatively lower \( C_{\text{ratio}} \) values in the region where \( N < \text{Sat} \). This indicates that a lower \( \omega_{\min} \) increases the potential for maximizing cost efficiency in the initial grouping process. Such a trend is reasonably understood in the context of grouping strategies being highly influenced by initial manufacturing costs.

\newpage
\subsection{Summary of the feasibility of modular design} 
\begin{figure*}[b!]
\centering
 \includegraphics[width=1\textwidth]{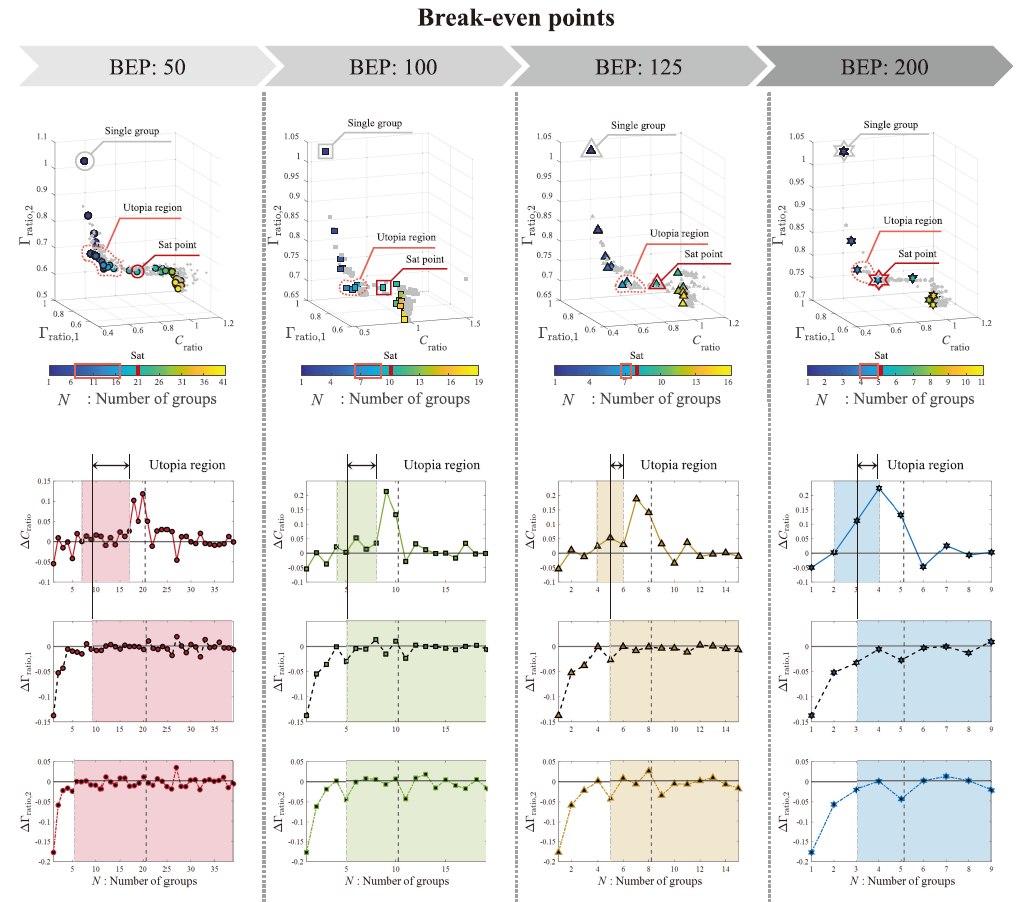}
\caption{Pareto front analysis under varying BEP conditions. The colormap highlights Pareto-optimal solutions for each BEP, indicating the trade-off between cost efficiency (\( C_{\text{ratio}} \)) and payload torque deviations (\( \Gamma_{\text{ratio},1}, \Gamma_{\text{ratio},2} \)) across different \( N \).}
\label{CH6_fig_result4}
\vspace{-9pt}
\end{figure*}

Achieving an optimal balance between manufacturing cost and performance is a critical challenge in large-scale engineering design scenarios. The proposed modular design strategy aims to enhance cost efficiency while maintaining design performance across various manufacturing conditions. To evaluate the feasibility, a Pareto analysis was performed on the three normalized objective functions (\( C_{\text{ratio},N}, \Gamma_{\text{ratio},1,N}, \Gamma_{\text{ratio},2,N} \)) using Eq.~\eqref{CH6_eq_origin2}. The results are presented in Fig.~\ref{CH6_fig_result4}, which visualizes the Pareto solutions selected under different BEP conditions through a colormap representation. This figure illustrates the trade-off relationship between cost (\( C_{\text{ratio},N} \)) and performance metrics (\( \Gamma_{\text{ratio},1,N}, \Gamma_{\text{ratio},2,N} \)), showing how these factors interact under varying manufacturing constraints.

As observed in Fig.~\ref{CH6_fig_result1}, the trade-off between the normalized objective functions becomes more apparent as the number of groups (\( N \)) increases. While \( C_{\text{ratio}} \) converges to 1 beyond Sat, \( \Gamma_{\text{ratio},1} \) and \( \Gamma_{\text{ratio},2} \) exhibit a gradual decrease. This indicates that although an increase in \( N \) leads to inefficiencies in cost, it reduces performance discrepancies among a large set of distinct optimal designs. This outcome can be reasonably interpreted as a result of finer-resolution grouping, which can be expressed as follows:


\begin{equation}\label{CH6_result4_conserve}
\lim_{N \to M} \frac{C(\mathbf{G}_N)}{C_{\text{origin}}} = 1, \quad 
\lim_{N \to M} \frac{\Delta\Gamma_1(\mathbf{G}_N)}{\Gamma_{\text{origin,1}}} = 0, \quad 
\lim_{N \to M} \frac{\Delta\Gamma_2(\mathbf{G}_N)}{\Gamma_{\text{origin,1}}} = 0.
\end{equation}

Equation~\eqref{CH6_result4_conserve} demonstrates that as the number of groups (\( N \)) approaches the total number of design candidates (\( M \)), the cost (\( C_{\text{ratio}} \)) converges to 1, while the two performance metrics (\( \Gamma_{\text{ratio},1} \), \( \Gamma_{\text{ratio},2} \)) approach 0. This indicates that when \( N \) reaches \( M \), every design candidate is produced individually, eliminating the cost advantages gained through modular design. Therefore, selecting an appropriate number of groups (\( N \)) is crucial in modular design strategies. 

Notably, near Sat, where \( N \) is determined by BEP, \( C_{\text{ratio}} \) exhibits a sharp increase (Fig.~\ref{CH6_fig_result1} (a) and Fig.~\ref{CH6_fig_result4}), whereas the variations in \( \Gamma_{\text{ratio},1} \) and \( \Gamma_{\text{ratio},2} \) remain relatively small (Fig.~\ref{CH6_fig_result1} (b) and Fig.~\ref{CH6_fig_result4}). This phenomenon reflects a situation in which cost (\( C_{\text{ratio}} \)) escalates rapidly despite minimal improvements in payload torque deviation. These findings suggest that the number of groups (\( N \)) near Sat must be carefully determined to balance cost efficiency and performance. In particular, in large-scale design problems, the Sat region is a critical decision point for optimizing the trade-off between cost savings and performance improvements when structuring the design layout.


The Utopia region can be defined as the area encompassing the Pareto solutions where the trade-off relationship is most pronounced and the objective functions exhibit saturation and convergence. To analyze this more clearly, the variation in the objective functions of the Pareto solution closest to the origin in terms of Euclidean distance was examined as the number of groups (\( N \)) increased, as shown in the lower section of Fig.~\ref{CH6_fig_result4}. 

In this analysis, the variation for each group number (\( N \)) was computed using a finite difference operation between adjacent group numbers without the need for separate regression analysis or approximation processes. Specifically, the variation terms \( \Delta C_{\text{ratio}}, \Delta \Gamma_{\text{ratio},1}, \Delta \Gamma_{\text{ratio},2} \) represent the difference when the group number increases from \( N \) to \( N+1 \). Consequently, the selection range of the Utopia region is determined based on the interval where these variations exhibit the most significant decline. However, due to the nature of the difference operation, the length of the computed variation vector is one less than the original data, requiring a correction of +1 in the analyzed range to determine the actual Utopia region. 

In Fig.~\ref{CH6_fig_result4}, the valid group number range near Sat for each BEP condition is indicated by a dashed line, encompassing the Utopia region that reflects the balance between cost (\( C_{\text{ratio}} \)) and performance metrics (\( \Gamma_{\text{ratio},1}, \Gamma_{\text{ratio},2} \)). The results indicate that for BEP = 50, the valid group number range was from 6 to 17. However, as BEP increased, the available range of group numbers gradually decreased. Specifically, for BEP = 100, the range was from 6 to 8; for BEP = 125, it was from 6 to 7; and for BEP = 200, the valid group range was significantly reduced to 4 to 5. This trend suggests that the potential for maintaining cost efficiency through grouping is considerably reduced as BEP increases.

Ultimately, the selection of the number of groups (\( N \)) near Sat plays a critical role in balancing \( C_{\text{ratio}}, \Gamma_{\text{ratio},1}, \Gamma_{\text{ratio},2} \). The results indicate that as BEP increases, the cost reduction effect achievable through grouping gradually diminishes, and under high BEP conditions, the viable group configurations become extremely limited. Consequently, the number of selectable groups is likely to be reduced to only one or two, and beyond a certain BEP threshold, the practical feasibility of grouping may be almost entirely lost, highlighting a structural limitation in the modular design strategy.



\newpage
\section{Conclusion}\label{sec5}
This study proposed a modular design framework that enhances cost efficiency and sustainability in large-scale engineering applications. The modular design minimizes material waste and optimizes resource utilization by enabling large-scale production and realizing cost reduction through economies of scale. This approach enhances manufacturing efficiency and contributes to sustainable practices by reducing excess material consumption and improving production consistency. However, applying modularization to mechanism design presents unique challenges due to the inherent trade-offs between standardization and performance optimization. A cost-efficient and performance-balanced modular design strategy was developed to address this, specifically tailored for large-scale design scenarios. Unlike traditional engineering design approaches that primarily optimize individual design candidates using surrogate or generative models, this framework systematically integrates modular principles to overcome the challenge of achieving economies of scale.

As an application of modular mechanism design, the quasi-serial manipulator large-scale design scenario problem was selected. The proposed modular design strategy formulates the problem as a partitioning and grouping optimization, determining the optimal number of groups and the degree of standardization within each group. By defining the break-even point (BEP) and cost curve, the approach assumes that grouping designs beyond the BEP threshold would lead to cost advantages. However, group standardization introduces trade-offs, as performance deviations from individually optimized designs may occur. To minimize this deviation while maintaining cost efficiency, the optimal design layout was identified by systematically balancing performance consistency and cost reduction.

The optimization framework was formulated as a multi-objective problem using normalized objective functions (\( C_{\text{ratio}}, \Gamma_{\text{ratio},1}, \Gamma_{\text{ratio},2} \)). By incrementally increasing the number of groups and evaluating corresponding objective metrics, saturation points in manufacturing costs were identified, and the optimal modular design layout was selected within the utopia region where performance deviation was minimized. The analysis of various manufacturing conditions revealed that the feasible range for cost-effective grouping rapidly diminished as the BEP increased. Near the saturation point (Sat), a sharp increase in cost (\( C_{\text{ratio}} \)) was observed, while performance metrics exhibited minimal variation, allowing for the identification of an optimal cost-performance balance. This finding underscores the necessity of clearly defining manufacturing cost parameters in the early design stage to determine the most effective modular design layout. Excessively high BEP values or unfavorable initial manufacturing costs may diminish the economic viability of the modular design strategy, making early-stage cost analysis a critical decision factor.

The primary contribution of this study is developing an efficient modular design strategy for large-scale engineering applications, integrating surrogate model-driven optimization with modular mechanism design problems. Unlike conventional approaches that focus on optimizing individual designs, the proposed framework systematically structures multiple optimal solutions within a shared platform, balancing standardization and performance consistency. Leveraging a surrogate-based optimization method efficiently captures complex performance trade-offs while minimizing computational costs. A production cost is also considered to identify economically feasible modular configurations based on break-even points and cost parameters. While demonstrated in manipulator design, this framework applies to a wide range of modular engineering systems, offering a scalable approach to structuring design variability in cost-sensitive manufacturing environments.


\bmhead{Acknowledgments}
This work was supported by the Ministry of Science and ICT of Korea grant (No.2022-0-00969, No.2022-0-00986, and GTL24031-000) and the Ministry of Trade, Industry \& Energy (RS-2024-00410810).

\section*{Declarations}
\bmhead{Conflict of interest}
The authors declare that they have no known competing financial interests or personal relationships that could influence the work reported herein.

\bmhead{Replication of results}
The code and data are available from the corresponding author on reasonable request.

\bibliography{sn-bibliography}

\end{document}